# Clustering Algorithms for the Centralized and Local Models[*]

Kobbi Nissim[†]    Uri Stemmer[‡]

July 15, 2017

## Abstract

We revisit the problem of finding a minimum enclosing ball with differential privacy: Given a set of $n$ points in the Euclidean space $\mathbb{R}^d$ and an integer $t \leq n$, the goal is to find a ball of the smallest radius $r_{opt}$ enclosing at least $t$ input points. The problem is motivated by its various applications to differential privacy, including the sample and aggregate technique, private data exploration, and clustering [20, 21, 15].

Without privacy concerns, minimum enclosing ball has a polynomial time approximation scheme (PTAS), which computes a ball of radius almost $r_{opt}$ (the problem is NP-hard to solve exactly). In contrast, under differential privacy, until this work, only a $O(\sqrt{\log n})$-approximation algorithm was known.

We provide new constructions of differentially private algorithms for minimum enclosing ball achieving constant factor approximation to $r_{opt}$ both in the centralized model (where a trusted curator collects the sensitive information and analyzes it with differential privacy) and in the local model (where each respondent randomizes her answers to the data curator to protect her privacy).

We demonstrate how to use our algorithms as a building block for approximating $k$-means in both models.

[*]Research supported by NSF grant No. 1565387.
[†]Dept. of Computer Science, Georgetown University and Center for Research on Computation and Society (CRCS), Harvard University. kobbi.nissim@georgetown.edu.
[‡]Center for Research on Computation and Society (CRCS), Harvard University. u@uri.co.il.

# 1 Introduction

We revisit the problem of finding a minimum enclosing ball with differential privacy and provide efficient algorithms for both the centralized curator model and the local model. The problem is motivated by its various applications to differential privacy: it can be used as the aggregation function in the sample and aggregate framework [20, 21] as well as serve as a building block in private data exploration, and clustering [15]. Given a set of $n$ points in the Euclidean space $\mathbb{R}^d$ and an integer $t \leq n$, the goal is to find a ball of smallest radius $r_{opt}$ enclosing at least $t$ input points.

**Definition 1.1** (revised in Definition 2.1). *An instance of the* 1-cluster *problem is a collection of $n$ points in the $d$-dimensional Euclidean space and a target parameter $t \leq n$. Let $r_{opt}$ be the radius of the smallest ball containing at least $t$ of the points. A $(\Delta, w)$-approximate solution for the instance consists of a ball of radius at most $w \cdot r_{opt}$ containing at least $t - \Delta$ of the points.*

Our recent work with Vadhan [21] provided an efficient differentially private $(\Delta, w)$-approximation algorithm for 1-cluster for $t > O(\sqrt{d} \log n/\epsilon)$ with $\Delta = O(\log n/\epsilon)$ and $w = O(\sqrt{\log n})$ (for clarity we omit the dependency on some of the parameters). In words, this algorithm identifies with differential privacy a ball of radius $O(r_{opt} \cdot \sqrt{\log n})$ containing at least $t - O(\log n/\epsilon)$ points, provided that $t > O(\sqrt{d} \log n/\epsilon)$. This algorithm works in the centralized curator model, where all data is collected and processed by a trusted curator.

**A new construction based on locality sensitive hashing:** Our first contribution is a new algorithm for the centralized curator model. Our new algorithm `LSH-GoodCenter` has two main components. First, a family of *locality sensitive hash functions* [17] is utilized to identify a small number of disjoint subsets of the input points, such that (at least) one of these subsets is contained in an (approximately) minimal enclosing ball. The identification is indirect in that it provides a predicate that evaluates to one on the "identified" input points and zero on all other input points. Then, these identified points are *averaged with differential privacy* to locate a point within their enclosing ball. We give a construction following these lines for solving 1-cluster in the $d$-dimensional Euclidean space. The resulting algorithm provides constant approximation to $r_{opt}$ (i.e., $w = O(1)$). Furthermore, the construction lends itself to metrics other than the Euclidean metric, provided that locality preserving hashing exist and that some "averaging" can be performed in the metric with differential privacy. Table 1 summarizes this result in comparison with previous algorithms for the problem.

## 1.1 Local algorithms

Local algorithms are a sub-class of differentially private algorithms where each individual respondent randomizes her own data to guarantee her own privacy. Hence, respondents in a local algorithm do not need to trust a centralized curator with their data. The accuracy of local algorithms is, generally, lower than what is achievable with a trusted curator, however, where data exists in abundance, local algorithms can provide a practical alternative for computing over sensitive data with minimal trust assumptions. Indeed, recent industrial implementation of differential privacy by Google [12] and Apple [24] are reported to utilize local algorithms.

**A local algorithm for 1-cluster:** Our second contribution is an algorithm for 1-cluster in the local model. The main part of the construction, Algorithm `LDP-GoodCenter`, combines locality sensitive hashing with the recent optimal heavy hitters local algorithm of Bassily et al. [2] to identify a subset of input points that fall within an approximately minimal enclosing ball. We then approximate the average of the identified points using a new (and simple) construction for approximating averages of points under local differential privacy. Importantly, the respondents running time and communication complexity is only polylogarithmic in the number of participants, and the running time for the data curator is slightly super linear.

|  | Needed cluster size – $t$<br>Additive loss in cluster size – $\Delta$ | Approximation factor<br>in radius – $w$ |
|---|---|---|
| [20]<br>(centralized model) | $t \geq \max\left\{0.51n, O(\frac{d^2}{\epsilon^2}\log^2|X|)\right\}$<br>$\Delta = \frac{1}{\epsilon} \cdot \log\log(d|X|)$ | $w = O(\sqrt{d}/\epsilon)$ |
| [21]<br>(centralized model) | $t \geq \frac{\sqrt{d}}{\epsilon} \cdot \log^{1.5}\left(\frac{1}{\delta}\right) \cdot 2^{O(\log^*(|X|d))}$<br>$\Delta = \frac{1}{\epsilon} \cdot \log\left(\frac{1}{\delta}\right) \cdot 2^{O(\log^*(|X|d))}$ | $w = O(\sqrt{\log n})$ |
| This work<br>(centralized model) | $t \geq \frac{n^{0.1} \cdot \sqrt{d}}{\epsilon} \cdot \log^{1.5}\left(\frac{1}{\delta}\right) \cdot 2^{O(\log^*(|X|d))}$<br>$\Delta = \frac{n^{0.1}}{\epsilon} \cdot \log\left(\frac{1}{\delta}\right) \cdot 2^{O(\log^*(|X|d))}$ | $w = O(1)$ |
| This work<br>(local model) | $t \geq O\left(\frac{1}{\epsilon} \cdot n^{0.67} \cdot d^{1/3} \cdot \log(n|X|)\right)$<br>$\Delta = O\left(\frac{1}{\epsilon} \cdot n^{0.67} \cdot \log(n|X|)\right)$ | $w = O(1)$ |

Table 1: Algorithms for 1-cluster

## 1.2 Application to clustering

Feldman et al. [15] showed that an algorithm for privately solving the 1-cluster problem can be used for privately computing approximations to the $k$-means of the input, by iterating the algorithm for (roughly) $k' = k \cdot \log(n)$ times and finding $k'$ balls that cover most of the data points (the $k'$ centers can then be post-processed to find $k$ centers that approximate the $k$-means of the data points). Our new algorithm for the 1-cluster problem immediately yields new bounds on the achievable error for privately approximating the $k$-means.

**$k$-mean clustering in the local model:** We show that the construction of [15] can be implemented in the local model. Together with our local algorithm for the 1-cluster problem, this results in the first local differentially private algorithm that computes a provable approximation of the $k$-means clustering problem. As with our local algorithm for the 1-cluster problem, the running time and the communication complexity of the users is only polylogarithmic in the number of participants.

## 2 Preliminaries

**Notation.** Throughout the paper, we use $X$ to denote a finite totally ordered data domain, and use $X^d$ for the corresponding $d$-dimensional domain. We identify $X^d$ with points in the real $d$- dimensional unit cube, quantized with grid step $1/(|X|-1)$. Datasets are (ordered) collections of elements from some data universe $U$ (e.g., $U = X$ or $U = X^d$ or $U = \mathbb{R}^d$). Two datasets $S, S' \in U^n$ are called *neighboring* if they differ on at most one entry.

## 2.1 1-cluster

The reason for our explicit embedding of the finite ordered domain $X$ in $\mathbb{R}$ is that certain computations are impossible for differential privacy when the input domain is the unit interval $[0, 1]$. An example is the *Interior Point* problem [6] where the task is, given $n$ points in the unit interval $x_1, x_2, \ldots, x_n$, to output a value in the range $[\min(x_1, \ldots, x_n), \max(x_1, \ldots, x_n)]$. This impossibility result carries over to the 1-cluster problem (through a reduction from the interior point problem) [21]. With this fact in mind, we now redefine 1-cluster:

**Definition 2.1.** *A* 1-cluster *problem* $(X^d, n, t)$ *consists of a d-dimensional domain* $X^d$ *and parameters* $n \geq t$. *We say that algorithm* $\mathcal{M}$ *solves* $(X^d, n, t)$ *with parameters* $(\Delta, w, \beta)$ *if for every input database* $S \in (X^d)^n$ *it outputs, with probability at least* $1 - \beta$, *a center* $c$ *and a radius* $r$ *such that (i) the ball of radius* $r$ *around* $c$ *contains at least* $t - \Delta$ *points from* $S$; *and (ii)* $r \leq w \cdot r_{opt}$, *where* $r_{opt}$ *is the radius of the smallest ball in* $X^d$ *containing at least* $t$ *points from* $S$.

## 2.2 Preliminaries from differential privacy

Consider a database, where each entry contains information pertaining to an individual. An algorithm operating on databases is said to preserve differential privacy if a change of a single record of the database does not significantly change the output distribution of the algorithm. Intuitively, this means that individual information is protected: whatever is learned about an individual could also be learned with her data arbitrarily modified (or without her data).

**Definition 2.2** (Differential Privacy [9])**.** *A randomized algorithm* $M : U^n \to Y$ *is* $(\epsilon, \delta)$ *differentially private if for every two neighboring datasets* $S, S' \in U^n$ *and every* $T \subseteq Y$ *we have* $\Pr[M(S) \in T] \leq e^\epsilon \cdot \Pr[M(S') \in T] + \delta$, *where the probability is over the randomness of* $M$.

### 2.2.1 The Laplace and Gaussian Mechanisms

The most basic constructions of differentially private algorithms are via the Laplace and Gaussian mechanisms as follows.

**Definition 2.3** ($L_p$-Sensitivity)**.** *A function* $f$ *mapping databases to* $\mathbb{R}^d$ *has* $L_p$-sensitivity $k$ *if* $\|f(S) - f(S')\|_p \leq k$ *for all neighboring* $S, S'$.

**Theorem 2.4** (Laplace mechanism [9])**.** *A random variable is distributed as* $\mathrm{Lap}(\lambda)$ *if its probability density function is* $f(y) = \frac{1}{2\lambda} \exp(-\frac{|y|}{\lambda})$. *Let* $\epsilon > 0$, *and assume* $f : U^* \to \mathbb{R}^d$ *has* $L_1$-sensitivity $k$. *The mechanism that on input* $S \in U^*$ *outputs* $f(S) + \left(\mathrm{Lap}(\frac{k}{\epsilon})\right)^d$ *is* $(\epsilon, 0)$-differentially private.

**Theorem 2.5** (Gaussian Mechanism [7])**.** *Let* $\epsilon, \delta \in (0, 1)$, *and assume* $f : U^* \to \mathbb{R}^d$ *has* $L_2$-sensitivity $k$. *Let* $\sigma \geq \frac{k}{\epsilon}\sqrt{2\ln(1.25/\delta)}$. *The mechanism that on input* $S \in U^*$ *outputs* $f(S) + \left(\mathcal{N}(0, \sigma^2)\right)^d$ *is* $(\epsilon, \delta)$-differentially private.

### 2.2.2 Stability based histogram [8, 23, 3]

Given a database $S \in U^*$, consider the task of choosing a "good" solution out of a possible solution set $F$ where "goodness" is measured by a *quality function* $q : U^* \times F \to \mathbb{N}$ assigning "scores" to solutions from $F$ (w.r.t. the given database $S$). One of the constructions in differential privacy – the exponential mechanism [19] – can privately identify an approximately optimal solution $f \in F$ provided that $q$ has low-sensitivity and that $|S| \gtrsim \log |F|$.

By focusing on cases where the number of possible solutions with "high" scores is limited, it is possible to relax the requirement that $|S| \gtrsim \log |F|$, using what has come to be known as *stability based techniques*. In this work we use stability based techniques for the following task: Given a dataset $S \in U^n$ and a partition $P$ of $U$, find a set $p \in P$ containing (approximately) maximum number of elements of $S$.

**Theorem 2.6** ([3, 5, 25]). *Fix $\beta, \epsilon, \delta, n$, and let $t \geq \frac{12}{\epsilon} \ln(\frac{n}{\beta \delta})$. There exists an $(\epsilon, \delta)$-differentially private algorithm that given $S \in U^n$ returns a collection $L \subset P$ of sets in the partition $P$ such that $|\{x \in S : x \in p\}| \geq t/4$ for every $p \in L$ (hence, there are at most $4n/t$ sets in the collection $L$); and, with probability at least $(1 - \beta)$, the collection $L$ contains every set $p \in P$ such that $|\{x \in S : x \in p\}| \geq t$.*

### 2.2.3 The sparse vector technique [10]

Consider a sequence of low sensitivity functions $f_1, f_2, \ldots, f_k$, which are given (one by one) to a data curator (holding a database $S$). Algorithm `AboveThreshold` by Dwork et al. [10] privately identifies the first query $f_i$ whose value $f_i(S)$ is greater than some threshold $t$:

**Theorem 2.7** (Algorithm `AboveThreshold` [10]). *There exists an $(\epsilon, 0)$-differentially private algorithm $\mathcal{A}$ such that for $k$ rounds, after receiving a sensitivity-1 query $f_i : U^* \to \mathbb{R}$, algorithm $\mathcal{A}$ either outputs $\top$ and halts, or outputs $\bot$ and waits for the next round. If $\mathcal{A}$ was executed with a database $S \in U^*$ and a threshold parameter $t$, then the following holds with probability $(1-\beta)$: (i) If a query $f_i$ was answered by $\top$ then $f_i(S) \geq t - \frac{8}{\epsilon} \log(2k/\beta)$; (ii) If a query $f_i$ was answered by $\bot$ then $f_i(S) \leq t + \frac{8}{\epsilon} \log(2k/\beta)$.*

### 2.2.4 Composition theorems

We construct algorithms that use several differentially private mechanisms as subroutines, and analyze the overall privacy using the following composition theorems:

**Theorem 2.8** (Basic composition [7, 8]). *A mechanism that permits $k$ adaptive interactions with $(\epsilon, \delta)$-differentially private mechanisms (and does not access the database otherwise) is $(k\epsilon, k\delta)$-differentially private.*

**Theorem 2.9** (Advanced composition [11]). *Let $\epsilon, \delta, \delta' > 0$. A mechanism that permits $k$ adaptive interactions with $(\epsilon, \delta)$-differentially private mechanisms (and does not access the database otherwise) is $(\epsilon', k\delta + \delta')$-differentially private, for $\epsilon' = 2k\epsilon^2 + \epsilon \sqrt{2k \ln(1/\delta')}$.*

## 3 Our Algorithms

Let $X^d$ be a a finite subset of the Euclidean space $\mathbb{R}^d$. We consider the following problem under differential privacy:

**Definition 3.1** (The problem of minimum enclosing ball with $t$ points). *Given a set $S$ of $n$ points in $X^d$ and a parameter $t \leq n$, find a ball of minimal radius $r_{opt}$ enclosing at least $t$ input points.*

In this section we give an informal description of our algorithms for this problem and highlight some of the ideas behind the constructions. Any informalities made herein will be removed in the sections that follow. Consider the following relaxed variant of the above problem:

**Definition 3.2** (Minimum enclosing ball – promise version). *Given a set $S$ of $n$ points in $X^d$, an integer $t \leq n$, and a radius $r \in \mathbb{R}$ s.t. there exists a ball in $X^d$ containing at least $t$ points from $S$, the goal is to find a ball of radius $r$ enclosing at least $t$ input points.*

That is, in addition to the input set $S$ and the parameter $t$, in this promise problem we are also given a target radius $r$, and our goal is to identify a ball of that radius that contains $t$ of the input points, under the promise that such a ball exists.

As was observed in [21], it suffices to solve the promise problem. To see this, fix a database $S \in (X^d)^n$, and let $r_{opt}$ denote the smallest radius s.t. there is a ball of that radius containing $t$ points from $S$. Assume that we have an algorithm $\mathcal{A}$ for the promise problem. Now, for every possible choice for a radius $r$, consider applying $\mathcal{A}$ on the database $S$ with the radius $r$ to obtain a center $c_r$. As $\mathcal{A}$ solves the promise problem, for every choice for $r$ s.t. $r \geq r_{opt}$, we will have that the ball of radius $r$ around the obtained center $c_r$ contains at least $t$ points from $S$. Using standard tools from differential privacy (e.g., the Laplace mechanism), we can now privately estimate, for every $r$, the number of input points contained in the ball of radius $r$ around $c_r$. Afterwards we can return the center $c_{r^*}$ and the radius $r^*$ such that $r^*$ is the minimal radius for which (our estimation for) the number of points in the corresponding ball is at least $\approx t$. In fact, as we are only aiming for a solution with an approximated radius $r$, it suffices to only consider radiuses in powers of 2. As our domain $X^d$ is finite, there are at most $\log(d|X|)$ such choices, so this will introduce error at most $\text{polylog}(d|X|)$ in our estimations (meaning that we will identify a ball containing $\approx t - \text{polylog}(d|X|)$ input points). Furthermore, a binary search for the radius $r^*$ (out of all the possible powers of 2) would reduce the error to $\text{polyloglog}(d|X|)$. Using the privacy-preserving recursion on binary search of [3] it is possible to reduce the incurred error even further, to $2^{O(\log^*(d|X|))}$. In the rest of this section we focus on solving (or rather approximating) the promise problem under differential privacy. To that end, let us take an even closer look at the problem, and define the following restricted variant:

**Definition 3.3** (Promise problem, restricted to $t = n$). *Given a set $S$ of $t$ points in $X^d$ and a radius $r \in \mathbb{R}$ s.t. there exists a ball in $X^d$ containing all $t$ input points, the goal is to locate a ball of radius $r$ enclosing all $t$ input points.*

Intuitively, this makes the problem easier since we no longer have to identify which of the $n$ points are in the 1-cluster (i.e., which $t$ points can be enclosed in a ball of radius $r$). Indeed, without privacy concerns, this restricted variant of the problem is easier, and algorithms for the restricted problem are sometimes used as a subroutine for approximating the general problem.[1] Similarly, Nissim et al. [21] showed that in order to obtain a *private* approximation for the unrestricted problem (Definition 3.2) it suffices to *privately* approximate the restricted problem (Definition 3.3).

---

[1] Computing an exact solution to the problem of a minimum ball enclosing $t$ points is NP-hard, even for the case where $t = n$. However, better approximation algorithms exist for the restricted case where $t = n$. The restricted version has a FPTAS while the general problem does not (assuming that $P \neq NP$). See, e.g., [22].

However, their reduction introduced a factor of $\sqrt{\log n}$ in the radius of the computed ball. Our main technical contribution can be thought of as a new reduction from the problem of privately approximating a solution for the general problem to privately approximating a solution for the restricted problem while introducing only a constant blowup in the radius.

We first briefly describe the techniques of [21] for privately approximating the restricted problem: Assume that we are given a set $S$ of $t$ input points in $\mathbb{R}^d$, and a radius $r$ such that the set $S$ can be enclosed in a ball of radius $r$. Our goal is to locate a center $c$ such that a ball of radius $O(r)$ around $c$ contains *all* of $S$. The strategy here is conceptually simple – compute (an estimate for) the average of the points in $S$. Indeed, as $S$ is of diameter $2r$, a ball of radius $2r$ around this average contains all of $S$. Estimating the average of $S$ can be done privately using the Gaussian mechanism, with error proportional to $\frac{\sqrt{d}}{|S|}$ times the diameter of $S$, i.e., proportional to $\frac{\sqrt{d} \cdot r}{|S|}$. Hence, assuming that $t = |S| \gtrsim \sqrt{d}$, a ball of radius $O(r)$ around the (privately obtained) estimation for the average contains all of $S$.[2]

## 3.1 Our contributions

Our final task is to transform a private algorithm for the restricted promise problem into a private algorithm for the unrestricted promise problem, without incurring a blowup of $\sqrt{\log n}$ in the radius. As we next explain, this can be achieved using *locality sensitive hashing*. Informally, a locality sensitive hash function aims to maximize the probability of a collision for similar items, while minimizing the probability of collision for dissimilar items. Formally,

**Definition 3.4** ([17]). *Let $\mathcal{M}$ be a metric space, and let $r>0$, $c>1$, $0 \leq q < p \leq 1$. A family $\mathcal{H}$ of functions mapping $\mathcal{M}$ into domain $U$ is an $(r, cr, p, q)$ locality sensitive hashing family (LSH) if for all $x, y \in \mathcal{M}$ (i) $\Pr_{h \in_R \mathcal{H}}[h(x) = h(y)] \geq p$ if $d_\mathcal{M}(x, y) \leq r$; and (ii) $\Pr_{h \in_R \mathcal{H}}[h(x) = h(y)] \leq q$ if $d_\mathcal{M}(x, y) \geq cr$.*

Now, let $S \in \mathbb{R}^n$ be an input database and let $r \in \mathbb{R}$ be s.t. there exists a ball of radius $r$ that contains at least $t$ points from $S$. Let $P \subseteq S$ denote the guaranteed set of $t$ input points contained in a ball of radius $r$. Assume that we have an $(r, cr, p, q)$ locality sensitive hashing family $\mathcal{H}$ mapping $\mathbb{R}^d$ to some domain $U$, and sample a function $h \in_R \mathcal{H}$. Intuitively, if $q$ (the collision probability of dissimilar elements) is less than $n^{-2}$, then w.h.p. we will have that *all* dissimilar elements (i.e., at distance more than $cr$) are mapped into different hash values. On the other hand, as $P \subseteq S$ is of diameter $r$, we expect that $\approx p \cdot |P| = pt$ of the elements of $P$ will be mapped into the same hash value. In the following sections we show that this intuition is indeed correct, and that with noticeable probability the following two events occur (over the choice of $h \in \mathcal{H}$):

($E_1$) For every $x, y \in S$ s.t. $\|x - y\| \geq cr$ it holds that $h(x) \neq h(y)$; and,

($E_2$) There exists a hash value in $U$, denoted $u^*$, such that $|\{x \in P : h(x) = u^*\}| \geq \frac{p}{2} \cdot t$.

Event ($E_1$) states that if two points in $S$ are mapped into the same hash value, then these points are close. Event ($E_2$) states that there is a "heavy" hash value $u^* \in U$, such that "many" of the points in $P$ are mapped into $u^*$.

---

[2]For technical reasons, we cannot apply the Gaussian mechanism directly. Following [21], before using the Gaussian mechanism, we first privately compute a *box* of diameter $\approx r$ that contains all of $S$, and use it to bound the diameter of $S$.

Using standard stability based arguments (see, e.g., Theorem 2.6), we can now privately identify a list $L$ containing all such "heavy" hash values $u \in U$ (in particular, we will have that $u^* \in L$). Once we have the list $L$, we can use every $u \in L$ as a "filter" on $S$ that isolates clustered points: $S_u = \{x \in S : h(x) = u\}$. Observe that, by Event $E_1$, for every $u \in L$ we have that $S_u$ can be enclosed in a ball of radius $cr$. Hence, for every $u \in L$ we can use a private algorithm for the restricted promise problem on the database $S_u$ with the radius $cr$, to obtain a center $y_u$ s.t. a ball of radius $O(cr)$ around $y_u$ contains all of $S_u$.

So, we have generated a set of centers $Y$, such that there exists $y_{u^*} \in Y$ for which a ball of radius $O(cr)$ centered at $y_{u^*}$ contains all of $S_{u^*}$. As $S_{u^*}$ contains some of the points from the guaranteed cluster $P \subseteq S$ of diameter $2r$, we get that the ball of radius $O(cr + 2r) = O(cr)$ around $y_{u^*}$ contains *all* of $P$, and hence, contains at least $t$ points from $S$. All that remains is to privately identify a center $y \in Y$ s.t. the ball of radius $O(cr)$ around it contains $\approx t$ points from $S$. This can be done privately, e.g., using the sparse vector technique (see Theorem 2.7).

**Choosing the LSH parameters.** Recall that for the discussion above we needed that $q$ (the collision probability of dissimilar elements) is less than $n^{-2}$. In addition, in order to obtain a constant factor approximation in the radius, we want to have $c = O(1)$. Under those requirements, existing LSH constructions can achieve $p = n^{-b}$ for some small constant $b > 0$ (for example, see [1]). This, in turn, means that we need to have $t \gg n^{-b}$ in order for Event $E_2$ above to be meaningful.

We obtain the following theorem.

**Theorem 3.5.** *Let $n, t, \beta, \epsilon, \delta$ be s.t.*

$$t \geq O\left(\frac{n^{0.1} \cdot \sqrt{d}}{\epsilon} \log\left(\frac{1}{\beta}\right) \log\left(\frac{nd}{\beta\delta}\right) \sqrt{\log\left(\frac{1}{\beta\delta}\right)} \cdot 9^{\log^*(2|X|\sqrt{d})}\right).$$

*There exists an $(\epsilon, \delta)$-differentially private algorithm that solves the 1-cluster problem $(X^d, n, t)$ with parameters $(\Delta, w)$ and error probability $\beta$, where $w = O(1)$ and*

$$\Delta = O\left(\frac{n^{0.1}}{\epsilon} \log\left(\frac{1}{\beta\delta}\right) \log\left(\frac{1}{\beta}\right) \cdot 9^{\log^*(2|X|\sqrt{d})}\right).$$

In words, there exists an efficient $(\epsilon, \delta)$-differentially private algorithm that (ignoring logarithmic factors) is capable of identifying a ball of radius $O(r_{opt})$ containing $t - \tilde{O}(\frac{n^{0.1}}{\epsilon})$ points, provided that $t \geq \tilde{O}(n^{0.1} \cdot \sqrt{d}/\epsilon)$. The exponent 0.1 in the factor $n^{0.1}$ is arbitrary and can be reduced to any constant by appropriately choosing the LSH parameters. See Section 4 for the complete construction and analysis.

### 3.1.1  1-cluster in the local model

In Section 5 we show that many of the ideas discussed above have analogues for the *local model*, where each individual holds her private information locally, and only releases the outcomes of privacy-preserving computations on her data. While our construction for the local model is conceptually similar to the centralized construction, several modifications are required. These include introducing a new protocol for computing averages of points in $\mathbb{R}^d$ (satisfying local differential privacy), and a modified "filtering" usage of the locality sensitive hash function.

We obtain the following theorem, which is similar in spirit to Theorem 3.5, albeit with weaker guarantees.

**Theorem 3.6.** *Let $S = (x_1, \ldots, x_n) \in (X^d)^n$ be a database which is distributed across $n$ players (each holding one point in $X^d$). Let $t, \beta, \epsilon$ be s.t.*

$$t \geq O\left(\frac{1}{\epsilon} \cdot n^{0.67} \cdot d^{1/3} \cdot \log(dn|X|)\sqrt{\log(\frac{1}{\beta})}\right).$$

*There exists an algorithm, satisfying $\epsilon$-local differential privacy (LDP), that identifies a center $y$ and a radius $r$ s.t. with probability at least $1 - \beta$,*

1. *$r = O(r_{opt})$, where $r_{opt}$ is the radius of a smallest ball containing at least $t$ points from $S$.*

2. *The number of points from $S$ contained in the ball of radius $r$ around $y$ is at least $t - \Delta$, where*

$$\Delta = O\left(\frac{1}{\epsilon} \cdot n^{0.67} \cdot \sqrt{\log(d|X|)\log(\frac{1}{\beta})\log(\frac{n}{\beta})}\right).$$

Our algorithm uses three rounds of communication with the users. Constructing a non-interactive LDP algorithm for the 1-cluster problem (while maintaining constant error in the radius) remains an intriguing open problem.

## 4 Locating a small cluster – the centralized setting

We begin with an algorithm for the trusted curator model. As discussed in Section 3, we use a tool from [21] that enable us to focus on the task of *locating* a ball of radius $r$, under the assumption that $r$ is given to us and that $r$ is such that there exists a ball of radius $r$ enclosing at least $t$ of the input points.

### 4.1 Additional preliminaries

#### 4.1.1 Tool from prior work: Algorithm `GoodRadius` [21]

Let $S \subseteq X^n$ be a set of input points. Given a parameter $t$, consider the task of privately identifying a radius $r$ s.t. there exists a ball of radius $r$ containing $\approx t$ points from $S$, and furthermore, $r$ is not much bigger than the smallest such possible radius. Using privacy preserving recursion on binary search [3], Nissim et al. [21] obtained the following construction:

**Theorem 4.1** (Algorithm `GoodRadius`, [21]). *Let $S \in (X^d)^n$ be a database containing $n$ points from $X^d$ and let $t, \beta, \epsilon, \delta$ be parameters. There exists a $\mathrm{poly}(n, d, \log|X|)$-time $(\epsilon, \delta)$-differentially private algorithm that on input $S$ outputs a radius $r \in \mathbb{R}$ s.t. with probability at least $(1 - \beta)$:*

1. *There is a ball in $X^d$ of radius $r$ containing at least $t - O\left(\frac{1}{\epsilon}\log(\frac{1}{\beta\delta}) \cdot 9^{\log^*(|X|\cdot d)}\right)$ input points.*

2. *$r \leq 4 \cdot r_{opt}$ where $r_{opt}$ is the radius of the smallest ball in $X^d$ containing at least $t$ points from $S$.*

### 4.1.2 Random rotation

We also use the following technical lemma to argue that if a set of points $P$ is contained within a ball of radius $r$ in $\mathbb{R}^d$, then by randomly rotating the Euclidean space we get that (w.h.p.) $P$ is contained within an axis-aligned rectangle with side-length $\approx r/\sqrt{d}$.

**Lemma 4.2** (e.g., [26]). *Let $P \in (\mathbb{R}^d)^m$ be a set of $m$ points in the $d$ dimensional Euclidean space, and let $Z = (z_1, \ldots, z_d)$ be a random orthonormal basis for $\mathbb{R}^d$. Then,*

$$\Pr_Z \left[ \forall x, y \in P : \forall 1 \leq i \leq d : |\langle x - y, z_i \rangle| \leq 2\sqrt{\ln(dm/\beta)/d} \cdot \|x - y\|_2 \right] \geq 1 - \beta.$$

## 4.2 Algorithm `GoodCenter`

We are now ready to present algorithm `LSH-GoodCenter`. Given a radius $r$ computed by algorithm `GoodRadius` (Theorem 4.1), the algorithm privately locates a ball of radius $O(r)$ containing $\gtrsim t$ points.

The privacy properties of `LSH-GoodCenter` follow directly from composition (see Section 2.2.4). The analysis appears in Appendix A.

**Lemma 4.3.** *Algorithm `LSH-GoodCenter` preserves $(2\epsilon, 2\delta)$-differential privacy.*

We now proceed with the utility analysis of algorithm `LSH-GoodCenter`. We will assume the existence of a family $\mathcal{H}$ of $(r, cr, p{=}n^{-b}, q{=}n^{-2-a})$-sensitive hash functions mapping $\mathbb{R}^d$ to a universe $U$, for some constants $a > b$, $r > 0$, and $c > 1$.

**Lemma 4.4.** *Let `LSH-GoodCenter` be executed with the family $\mathcal{H}$ on a database $S$ containing $n$ points in $\mathbb{R}^d$, with parameters $\beta = n^{-a}/20$ and $t, \epsilon, \delta$ s.t.*

$$t \geq O\left(n^b \cdot \frac{\sqrt{d}}{\epsilon} \log\left(\frac{nd}{\delta}\right) \sqrt{\log(\frac{1}{\delta})}\right).$$

*If there exists a ball of radius $r$ in $\mathbb{R}^d$ containing at least $t$ points from $S$, then with probability at least $n^{-a}/4$, the output $\hat{y}$ in Step 7 is s.t. at least $t - O\left(\frac{1}{\epsilon} \log(n)\right)$ of the input points are contained in a ball of radius $2r(c+1)$ around $\hat{y}$.*

**Remark 4.5.** *We can think of $a$ and $b$ as small constants, e.g., $a = 0.2$ and $b = 0.1$. Hence, ignoring logarithmic factors, the above theorem only requires $t$ to be as big as $n^{0.1}\sqrt{d}/\epsilon$. Note that the algorithm only succeeds with small probability – namely, with probability $n^{-a}/4$. This can easily be amplified using repetitions (roughly $n^a = n^{0.2}$ repetitions).*

*Proof.* First recall that by the properties of the family $\mathcal{H}$, for every $x, y \in \mathbb{R}^d$ s.t. $\|x - y\| \geq cr$ we have that $\Pr_{h \in \mathcal{H}}[h(x) = h(y)] \leq q = n^{-2-a}$. Using the union bound we get

$$\Pr_{h \in_R \mathcal{H}}[h(x) \neq h(y) \text{ for all } x, y \in S \text{ s.t. } \|x - y\| \geq cr] \geq (1 - n^{-a}/2).$$

Let $P \subseteq S$ denote the guaranteed set of $t$ input points that are contained in a ball of radius $r$, and let $x \in P$ be an arbitrary point in $P$. By linearity of expectation, we have that

$$\mathop{\mathbb{E}}_{h \in \mathcal{H}} [|\{y \in P : h(y) \neq h(x)\}|] \leq t(1-p) = t(1 - n^{-b}).$$

Hence, by Markov's inequality,

$$\Pr_{h \in \mathcal{H}} \left[ |\{y \in P : h(y) \neq h(x)\}| \geq \frac{t(1-n^{-b})}{1-n^{-a}} \right] \leq 1 - n^{-a}.$$

So,

$$\Pr_{h \in \mathcal{H}} \left[ |\{y \in P : h(y) = h(x)\}| \geq t\left(1 - \frac{1-n^{-b}}{1-n^{-a}}\right) \right] \geq n^{-a}.$$

Simplifying, for large enough $n$ (specifically, for $n^{a-b} \geq 2$) we get

$$\Pr_{h \in \mathcal{H}} \left[ |\{y \in P : h(y) = h(x)\}| \geq \frac{t}{2} \cdot n^{-b} \right] \geq n^{-a}.$$

So far we have established that with probability at least $n^{-a}/2$ over the choice of $h \in \mathcal{H}$ in Step 1 the following events occur:

($E_1$) For every $x, y \in S$ s.t. $\|x - y\| \geq cr$ it holds that $h(x) \neq h(y)$; and,

($E_2$) There exists a hash value in $U$, denoted $u^*$, such that $|\{y \in P : h(y) = u^*\}| \geq \frac{t}{2} \cdot n^{-b}$.

Event ($E_1$) states that if two points in $S$ are mapped into the same hash value, then these points are close. Event ($E_2$) states that there is a "heavy" hash value $u^* \in U$, such that "many" of the points in $P$ are mapped into $u^*$. We proceed with the analysis assuming that these two events occur.

In Step 2 we use Theorem 2.6 to identify a list $L \in U^*$ that, w.p. $(1-\beta)$, contains all "heavy" hash values $u \in U$. In particular, with probability at least $(1-\beta)$ we have that $u^* \in L$. We proceed with the analysis assuming that this is the case.

In Step 3 we generate a random orthonormal basis $Z$. By Lemma 4.2, with probability at least $(1-\beta)$, for every $x, y \in S$ and for every $z_i \in Z$, we have that the projection of $(x-y)$ onto $z_i$ is of length at most $2\sqrt{\ln(dn/\beta)/d} \cdot \|x-y\|$. In particular, for every hash value $u \in L$ we have that the projection of $S_u \triangleq \{x \in S : h(x) = u\}$ onto every axis $z_i \in Z$ fits within an interval of length at most $p = 2rc\sqrt{\ln(dn/\beta)/d}$.

In Step 4 we use Theorem 2.6 again to identify, for every $u \in L$ and for every axis $z_i \in Z$, an interval of length $3p$ containing all of $u$. For our choice of $t$, all of these applications together succeed with probability at least $(1-\beta)$. This results, for every $u \in L$, in a box $B(u)$ of side length $3p$ and diameter $3p\sqrt{d}$ containing all of $S_u$.

For every $u \in L$ denote the average of the points in $S_u$ as $y_u$. In Step 5 we use the Gaussian mechanism to compute the noisy average $\hat{y}_u$ of $S_u$ for every $u \in L$. The noise magnitude reduces with the size of the sets $|S_u|$, and for our choice of $t$ (recall that $|S_u| \geq t/4$ for every $u \in L$), with probability at least $(1-\beta)$ we have that $\|y_u - \hat{y}_u\| \leq cr$ for every $u \in L$.

Recall that there exists a hash value $u^* \in L$ such that some of the points from the guaranteed cluster $P$ are mapped (by $h$) into $u^*$. Let $v \in P$ be such that $h(v) = u^*$. A ball of radius $2r$ around $v$ contains all of $P$, as $P$ is of radius $r$. In addition, by ($E_1$), a ball of radius $cr$ around $y_{u^*}$ contains all of $S_{u^*}$. As, $\|y_{u^*} - \hat{y}_{u^*}\| \leq cr$, we get that a ball of radius $2cr$ around $\hat{y}_{u^*}$ contains all of $S_{u^*}$, and in particular, contains $v$. Thus, a ball of radius $2cr + 2r$ around $\hat{y}_{u^*}$ contains all of $P$, that is, contains at least $t$ input points. Hence, by the properties of algorithm `AboveThreshold`, with probability at least $(1-\beta)$, the loop in Step 7 ends with algorithm `AboveThreshold` retuning $\top$, identifying a point $\hat{y}_u$ such that a ball of radius $2cr + 2r$ around it contains at least $t - \frac{65}{\epsilon} \log(2n/\beta)$ input points.

All in all, with probability at least $n^{-a}/2 - 5\beta = n^{-a}/4$ we have that algorithm `GoodCenter` outputs a point $\hat{y}$ such that a ball of radius $2r(c+1)$ around it contains at least $t - \frac{65}{\epsilon} \log(2n/\beta)$ input points.

$\square$

Theorem 3.5 now follows by combining Theorem 4.1 (algorithm `GoodRadius`) with Lemmas 4.3 and 4.4 (algorithm `LSH-GoodCenter`).

## 5 Locating a small cluster – the distributed setting

We begin by describing private computation in the local model where each individual holds her private information locally, and only releases the outcomes of privacy-preserving computations on her data. This is modeled by letting the algorithm access each entry $x_i$ in the input database $S = (x_1, \ldots, x_n) \in X^n$ separately, and only via differentially private *local randomizers*.

**Definition 5.1** (Local Randomizer, LR Oracle [18]). *A local randomizer $R : X \to W$ is a randomized algorithm that takes a database of size $n = 1$. Let $S = (x_1, \ldots, x_n) \in X^n$ be a database. An LR oracle $LR_S(\cdot, \cdot)$ gets an index $i \in [n]$ and a local randomizer $R$, and outputs a random value $w \in W$ chosen according to the distribution $R(x_i)$.*

**Definition 5.2** (Local differential privacy). *An algorithm satisfies $(\epsilon, \delta)$-local differential privacy (LDP) if it accesses the database $S$ only via the oracle $LR_S$ with the following restriction: for all possible executions of the algorithm and for all $i \in [n]$, if $LR_S(i, R_1), \ldots, LR_S(i, R_k)$ are the algorithm's invocations of $LR_S$ on index $i$, then the algorithm $\mathcal{B}(x) = (R_1(x), R_2(x), \ldots, R_k(x))$ is $(\epsilon, \delta)$-differentially private. Local algorithms that prepare all their queries to $LR_S$ before receiving any answers are called* noninteractive; *otherwise, they are* interactive.

### 5.1 Additional preliminaries

We now present additional preliminaries that enable our construction.

#### 5.1.1 Counting queries and histograms with local differential privacy

The most basic task we can apply under local differential privacy is a *counting query*. Consider a database $S \in X^n$ which is distributed across $n$ users (each holding one element from $X$). Given a predicate $y : X \to \{0, 1\}$, consider the task of estimating the number of users holding a value $x$ such that $y(x) = 1$. This can be solved privately with error proportional to $\frac{1}{\epsilon}\sqrt{n}$.

**Notation.** For a database $S \in X^n$ and a domain element $x \in X$, we use $f_S(x)$ to denote the duplicity of $x$ in $S$, i.e.,
$$f_S(x) = |\{x_i \in S : x_i = x\}|.$$

**Theorem 5.3** (e.g., [18]). *Let $\epsilon \leq 1$. Let $S = (x_1, \ldots, x_n) \in \{0,1\}^n$ be a database which is distributed across $n$ players (each holding one bit). There exists an algorithm satisfying $\epsilon$-LDP that returns an estimation $a \in \mathbb{R}$ s.t. with probability at least $1 - \beta$ we have that $|a - f_S(1)| \leq \frac{3}{\epsilon}\sqrt{n \cdot \ln(\frac{8}{\beta})}$.*

**Algorithm `LSH-GoodCenter`**

**Input:** Database $S \in \mathbb{R}^d$ containing $n$ points, radius $r$, target number of points in cluster $t$, failure probability $\beta$, privacy parameters $\epsilon, \delta$, and another parameter $c$ (reflecting multiplicative error in the radius of the found ball).

**Tool used:** A family $\mathcal{H}$ of $(r, c \cdot r, p, q)$-locality sensitive hash functions mapping $\mathbb{R}^d$ to a universe $U$.

1. Sample a hash function $h \in \mathcal{H}$ mapping $\mathbb{R}^d$ to $U$. For $u \in U$ define $S_u$ as the multiset containing all elements of $S$ that are mapped into $u$ by $h$, i.e., $S_u \triangleq \{x \in S : h(x) = u\}$.

2. Use algorithm from Theorem 2.6 with privacy parameters $(\frac{\epsilon}{4}, \frac{\delta}{4})$ to identify $L \subset U$ that, w.p. $(1-\beta)$, $L$ contains *every* hash value $u \in U$ s.t. $|S_u| \geq \frac{960\sqrt{d}}{\epsilon} \ln(\frac{nd}{\beta\delta})\sqrt{\ln(\frac{8}{\delta})}$. Moreover, for every $u \in L$ we have that $|S_u| \geq \frac{240\sqrt{d}}{\epsilon} \ln(\frac{nd}{\beta\delta})\sqrt{\ln(\frac{8}{\delta})}$.

    % W.h.p. $L$ contains a hash value $u$ s.t. a lot of the input points from the guaranteed cluster of radius $r$ are mapped into $u$.

3. Let $Z = (z_1, \ldots, z_d)$ be a random orthonormal basis of $\mathbb{R}^d$ and $p = 2rc\sqrt{\ln(\frac{dn}{\beta})/d}$. For each basis vector $z_i \in Z$, partition the axis in direction $z_i$ into intervals $\mathcal{I}_i = \{[j \cdot p, (j+1) \cdot p) : j \in \mathbb{Z}\}$.

4. For every hash value $u \in L$, identify (in an axis by axis manner) an axis aligned rectangle containing all of $S_u$. Specifically, for every $u \in L$:

   (a) For each basis vector $z_i \in Z$, use algorithm from Theorem 2.6 with privacy parameters $\left(\frac{\epsilon}{10\sqrt{d \ln(8/\delta)}}, \frac{\delta}{8d}\right)$ to choose an interval $I_i(u) \in \mathcal{I}_i$ containing a large number of points from $S_u$. Let $\hat{I}_i(u)$ denote that chosen interval after extending it by $p$ on each side (that is $\hat{I}_i(u)$ is of length $3p$).

   (b) Let $B(u)$ be the box in $\mathbb{R}^d$ whose projection on every basis vector $z_i \in Z$ is $\hat{I}_i(u)$. Define $S'_u = S_u \cap B(u)$.

      % Observe that we defined $S'_u = S_u \cap B(u)$ although we expect that $S_u \subseteq B(u)$. This will be useful in the privacy analysis, as we now have a deterministic bound on the diameter of $S'_u$.

5. For every hash value $u \in L$: Use the Gaussian mechanism with privacy parameters $(\frac{\epsilon}{4}, \frac{\delta}{4})$ to compute the noisy average of the points in $S'_u$. Denote the outcome as $\hat{y}_u$.

6. Instantiate algorithm `AboveThreshold` (Theorem 2.7) with database $S$, privacy parameter $\epsilon/4$, and threshold $t - \frac{33}{\epsilon}\log(2n/\beta)$.

7. For every hash value $u \in L$: Query algorithm `AboveThreshold` for the number of input points contained in a ball of radius $2r(c+1)$ around $\hat{y}_u$. If the answer is $\top$ then halt and return $\hat{y}_u$. Otherwise continue.

A slightly more complicated task is when instead of a binary predicate, we have a function $y : X \to U$ mapping domain elements to a (potentially) large set $U$. We use a tool from the recent work of Bassily et al. [2] on heavy hitters in the local model.

**Theorem 5.4** ([2]). *Let $\epsilon \leq 1$. Let $U$ be a data universe, and let $S = (u_1, \ldots, u_n) \in U^n$ be a database which is distributed across $n$ players (each holding one row). There exists an algorithm satisfying $\epsilon$-LDP that returns a set $L \subseteq (U \times \mathbb{R})$ of size at most $|L| \leq \sqrt{n}$ s.t. with probability $1 - \beta$ we have*

1. *For every $(u, a) \in L$ we have that $|a - f_S(u)| \leq O\left(\frac{1}{\epsilon}\sqrt{n \log(\frac{n}{\beta})}\right)$.*

2. *For every $u \in U$ s.t. $f_S(u) \geq O\left(\frac{1}{\epsilon}\sqrt{n \log(\frac{|X|}{\beta}) \log(\frac{1}{\beta})}\right)$, we have that $u$ appears in the set $L$.*

### 5.1.2 Concentration bounds

We will use the following variant of the Hoeffding bound for sampling without replacement.

**Theorem 5.5** (Hoeffdings inequality, sampling without replacement [16]). *Let $\mathcal{X} = (x_1, \ldots, x_N)$ be a finite population of $N$ points and $X_1, \ldots, X_n$ be a random sample drawn without replacement from $\mathcal{X}$. Let $a = \min_{i \in [N]}\{x_i\}$ and $b = \max_{i \in [N]}\{x_i\}$. Then, for all $\epsilon > 0$,*

$$\Pr\left[\left|\frac{1}{n}\sum_{i=1}^{n} X_i - \frac{1}{N}\sum_{i=1}^{N} x_i\right| \geq \epsilon\right] \leq 2 \exp\left(-\frac{2n\epsilon^2}{(b-a)^2}\right).$$

We also use the following tail bound on sums of $\lambda$-wise independent random variables.

**Lemma 5.6** ([4]). *Let $\lambda \geq 6$ be an even integer. Suppose $X_1, \ldots, X_n$ are $\lambda$-wise independent random variables taking values in $[0, 1]$. Let $X = X_1 + \cdots + X_n$ and $\mu = \mathbb{E}[X]$, and let $\alpha > 0$. Then,*

$$\Pr[|X - \mu| \geq \alpha] \leq \left(\frac{n\lambda}{\alpha^2}\right)^{\lambda/2}.$$

We also use a standard "Chernoff type" concentration bound for sums of i.i.d. samples from $\text{Lap}(b)$:

**Lemma 5.7.** *Suppose $X_1, \ldots, X_n$ are independent samples from $\text{Lap}(\frac{1}{\epsilon})$. That is, every $X_i$ has probability density function $f(x) = \frac{\epsilon}{2}\exp(-\epsilon|x|)$. Let $X = X_1 + \cdots + X_n$. For every $t \geq 0$ it holds that*

$$\Pr[|X| \geq t] \leq 6 \exp\left(-\frac{\epsilon \cdot t}{\sqrt{2n}}\right).$$

*Moreover, for every $0 \leq t < 2n/\epsilon$ we have that*

$$\Pr[|X| \geq t] \leq 2 \exp\left(-\frac{\epsilon^2 \cdot t^2}{4n}\right).$$

## 5.2 Average of vectors in $\mathbb{R}^d$

For our algorithms we will need a tool for privately computing averages of vectors in $\mathbb{R}^d$. Specifically, assume that there are $n$ users, where user $j$ is holding a point $x_j \in \mathbb{R}^d$. Moreover, assume that we have a bound $b \geq 0$ such that we are only interested in points $x_j \in \mathbb{R}^d$ s.t. every coordinate of $x_j$ is between 0 and $b$ (let us them "interesting points"). We would like to obtain an estimation (satisfying LDP) for the average of these interesting points. This can be achieved by having each user send a noisy version of its point. An estimation for the average of the interesting points can then be computed from the average of all the noisy points.

---
**Algorithm $\mathcal{R}$:** Local Randomizer for `LDP-AVG`

**Inputs:** $x \in \mathbb{R}$, privacy parameter $\epsilon$, and range parameter $b$.

1. If $x > b$ or $x < 0$ then set $x = 0$.
2. Let $z \leftarrow x + \text{Lap}(\frac{b}{\epsilon})$.
3. Return $z$.

---

---
**Algorithm `LDP-AVG`**

**Public parameters:** Random partition of $[n]$ into $d$ subsets $I_1, \ldots, I_d$.

**Setting:** Each player $j \in [n]$ holds a value $v_j \in \mathbb{R}^d \cup \{\bot\}$. Define $S = (v_1, \ldots, v_n)$.
  For $j \in [n]$ let $v'_j = v_j$ if every coordinate of $v_j$ is in $[0, b]$. Otherwise set $v'_j = \vec{0} \in \mathbb{R}^d$.
  For $j \in [n]$ let $x_j$ denote the $\ell^\text{th}$ coordinate of $v'_j$, where $\ell$ is s.t. $j \in I_\ell$.
  Define $\widetilde{S} = (x_1, \cdots, x_n)$.

**Oracle access:** LR Oracle to $S$ and $\widetilde{S}$.

1. For $j \in [n]$ let $z_j \leftarrow LR_{\widetilde{S}}(j, \mathcal{R})$ with parameters $\frac{\epsilon}{2}, b$.
2. For $\ell \in [d]$ define $y_\ell = \frac{n}{|I_\ell|} \sum_{j \in I_\ell} z_j$. Let $y = (y_1, \ldots, y_d) \in \mathbb{R}^d$.
3. Let $t$ denote the number of users $j$ holding a value $v_j \in \mathbb{R}^d$ s.t. every coordinate of $v_j$ is between 0 and $b$. Use Theorem 5.3 with parameter $\epsilon/2$ to obtain a noisy estimation $\hat{t}$ for $t$.
4. Return $\frac{1}{\hat{t}} \cdot y$.

---

**Observation 5.8.** *Algorithm `LDP-AVG` satisfies $\epsilon$-LDP.*

**Lemma 5.9.** *Let $b > 0$ be a parameter. Fix individual information $v_j \in \mathbb{R}^d \cup \{\bot\}$ for every user $j \in [n]$. Let $D \subseteq [n]$ denote the subset of users holding a value $v_j \neq \bot$ s.t. every coordinate of $v_j$ is between 0 and $b$, and denote $|D| = t$. Assume that $n \geq 8d \ln(\frac{8d}{\beta})$, and that $t \geq \frac{12}{\epsilon}\sqrt{n \log(\frac{32}{\beta})}$. Then, with probability at least $1 - \beta$, algorithm `LDP-AVG` returns a vector $a \in \mathbb{R}^d$ such that*

$$\left\| a - \frac{1}{t}\sum_{j \in D} v_j \right\|_2 \leq \frac{30bd}{t\epsilon}\sqrt{n \ln(\frac{32d}{\beta})}.$$

**Remark 5.10.** *For simplicity, in Lemma 5.9 we assumed that "interesting points" are points $x \in \mathbb{R}^d$ such that $x \in B$, where $B$ is a box with side-length $b$, aligned with the origin. This can easily be generalized to arbitrary axis-aligned boxes $B$ with side-length $b$ (that is, not necessarily located at the origin) by, e.g., shifting the axes accordingly.*

**Remark 5.11.** *Observe that the error in Lemma 5.9 decreases with $t$ (the number of "interesting points"). Specifically, for $t \gtrsim \sqrt{nd}$ we get that the error is proportional to the diameter of the box of interesting points, i.e., $\sqrt{d}b$.*

**Remark 5.12.** *Note that every user in the above protocol sends (on step 3 of algorithm $\mathcal{R}$) a real number $z$. In fact, it suffices to have each user send only $O(\log n)$ bits, without effecting the error guarantees of Lemma 5.9. To see this, observe that we can truncate $z$ to an interval of length $O\left(\frac{b}{\epsilon}\log(\frac{n}{\beta})\right)$. Indeed, by the properties of the Laplace distribution, w.h.p. this truncation has no effect on the computation. We can now discretize this interval with grid steps $O\left(\frac{b}{t\epsilon}\sqrt{\frac{d}{n}\log(\frac{d}{\beta})}\right)$, and have every user send the grid point closest to the point $z$, which can be represented using $O(\log n)$ bits. This introduces an error of at most $O\left(\frac{b}{t\epsilon}\sqrt{\frac{d}{n}\log(\frac{d}{\beta})}\right)$ per user, and hence error at most $O\left(\frac{b}{t\epsilon}\sqrt{dn\log(\frac{d}{\beta})}\right)$ per coordinate, and thus, error at most $O\left(\frac{bd}{t\epsilon}\sqrt{n\log(\frac{d}{\beta})}\right)$ in $L_2$ norm.*

*Proof of Lemma 5.9.* Our goal is to relate $y/\hat{t}$ (the output obtained on step 4) to $\frac{1}{t}\sum_{j \in D} v_j$. We will now show that w.h.p. every coordinate of $\left(\frac{y}{\hat{t}} - \frac{1}{t}\sum_{j \in D} v_j\right)$ is small.

First observe that by the Chernoff bound, assuming that $n \geq 8d\ln(\frac{8d}{\beta})$, with probability at least $1 - \frac{\beta}{4}$, for every $\ell \in [d]$ we have that $\frac{n}{2d} \leq |I_\ell| \leq \frac{2n}{d}$. We continue the analysis assuming that this is the case. Also, by Theorem 5.3, with probability at least $1 - \frac{\beta}{4}$ we have that $\hat{t}$ (computed on step 3) satisfies $|t - \hat{t}| \leq \frac{6}{\epsilon}\sqrt{n\log(\frac{32}{\beta})}$. In particular, assuming that $t \geq \frac{12}{\epsilon}\sqrt{n\log(\frac{32}{\beta})}$ we get that $\frac{t}{2} \leq \hat{t} \leq 2t$. We proceed with the analysis assuming that this is the case.

For $\ell \in [d]$, observe that $y_\ell$ (defined on Step 2 of the algorithm) can be expressed as

$$y_\ell = \frac{n}{|I_\ell|}\sum_{j \in I_\ell}(x_j + \eta_j) = \frac{n}{|I_\ell|}\sum_{j \in I_\ell \cap D} v_{j,\ell} + \frac{n}{|I_\ell|}\sum_{j \in I_\ell}\eta_j, \quad (1)$$

where $v_{j,\ell}$ is coordinate $\ell$ of $v_j$, and every $\eta_j$ is sampled from $\mathrm{Lap}(\frac{2b}{\epsilon})$. By the Hoeffding bound (sampling without replacement, Theorem 5.5), with probability at least $1 - \frac{\beta}{4}$ we have that

$$\left|\frac{n}{|I_\ell|}\sum_{j \in I_\ell \cap D} v_{j,\ell} - \sum_{j \in D} v_{j,\ell}\right| \leq bn\sqrt{\frac{1}{2|I_\ell|}\ln(8/\beta)} \leq b\sqrt{dn\ln(8/\beta)}. \quad (2)$$

Combining Equalities (1) and (2), and using the triangle inequality, we get

$$\left| \frac{y_\ell}{\hat{t}} - \frac{1}{t} \sum_{j \in D} v_{j,\ell} \right| \leq \frac{b\sqrt{dn \ln(8/\beta)}}{\hat{t}} + \left| \frac{n}{\hat{t}|I_\ell|} \sum_{j \in I_\ell} \eta_j \right| + \left| \left( \frac{1}{t} - \frac{1}{\hat{t}} \right) \sum_{j \in D} v_{j,\ell} \right|$$

$$\leq \frac{b\sqrt{dn \ln(8/\beta)}}{\hat{t}} + \left| \frac{n}{\hat{t}|I_\ell|} \sum_{j \in I_\ell} \eta_j \right| + \frac{|\hat{t} - t| \cdot b}{\hat{t}}$$

$$\leq \frac{2b}{t}\sqrt{dn \ln(8/\beta)} + \frac{4d}{t}\left| \sum_{j \in I_\ell} \eta_j \right| + \frac{12b}{t\epsilon}\sqrt{n \log(\frac{32}{\beta})}. \quad (3)$$

Finally, by Lemma 5.7 (tail bound for sum of i.i.d. samples from the Laplace distribution), with probability at least $1 - \frac{\beta}{4}$, for every $\ell \in [d]$ we have that $\left| \sum_{j \in I_\ell} \eta_j \right| \leq \frac{4b}{\epsilon}\sqrt{\frac{n}{d} \ln(\frac{8d}{\beta})}$. So,

$$(3) \leq \frac{2b}{t}\sqrt{dn \ln(8/\beta)} + \frac{16b}{t\epsilon}\sqrt{dn \ln(\frac{8d}{\beta})} + \frac{12b}{t\epsilon}\sqrt{n \log(\frac{32}{\beta})} \leq \frac{30b}{t\epsilon}\sqrt{dn \ln(\frac{32d}{\beta})}.$$

Overall, with probability at least $1 - \beta$ we get that every coordinate of the vector $\left( \frac{y}{\hat{t}} - \frac{1}{t} \sum_{j \in D} v_j \right)$ is at most $\frac{30b}{t\epsilon}\sqrt{dn \ln(\frac{32d}{\beta})}$ in absolute value. Hence, with probability at least $1 - \beta$ we have

$$\left\| \frac{y}{\hat{t}} - \frac{1}{t} \sum_{j \in D} v_j \right\|_2 \leq \frac{30bd}{t\epsilon}\sqrt{n \ln(\frac{32d}{\beta})}.$$

□

## 5.3 Algorithm `LDP-GoodCenter`

We are now ready to present our local algorithm for the 1-cluster problem. As we mentioned in the introduction, our algorithm can be used as a building block for privately approximating the $k$-means clustering problem, by iterating the algorithm and finding (roughly) $k$ balls that cover most of the data points. To that end, during the $i^{\text{th}}$ iteration we will need the ability to exclude from the computation the data points that were already covered in the previous iterations. This is handled by an additional input to the algorithm, a predicate $\sigma$, denoting a portion of the data universe to be excluded from the computation (meaning that if $\sigma(x) = 1$ then $x$ is to be excluded).

Our main statement for the 1-cluster problem under LDP is the following.

**Theorem 5.13.** *Fix a predicate $\sigma : X^d \to \{0, 1\}$. Let $S = (x_1, \ldots, x_n) \in (X^d)^n$ be a database which is distributed across $n$ players (each holding one point in $X^d$). Let $t, \beta, \epsilon$ be s.t.*

$$t \geq O\left( \frac{n^{0.51} \cdot d^{1/2}}{\epsilon} \log(dn|X|)\sqrt{\log(\frac{1}{\beta})} + \frac{n^{0.67} \cdot d^{1/3}}{\epsilon^{2/3}} \log(dn|X|) \left( \log(\frac{1}{\beta}) \right)^{1/3} \right).$$

*There exists an $\epsilon$-LDP algorithm that identifies a center $y$ and a radius $r$ s.t. with probability at least $1 - \beta$,*

1. $r = O(r_{opt})$, where $r_{opt}$ is the radius of a smallest ball containing at least $t$ points from $S_\sigma = \{x \in S : \sigma(x) \neq 1\}$.

2. The number of points from $S_\sigma$ contained in the ball of radius $r$ around $y$ is at least $t - \Delta$, where
$$\Delta = O\left(\frac{1}{\epsilon} \cdot \frac{n^{1.01}}{t^{1/2}} \cdot \sqrt{\log(d|X|)\log(\frac{1}{\beta})\log(\frac{n}{\beta})}\right).$$

The main part of our construction is algorithm `LDP-GoodCenter` which, assuming $t$ of the input points are contained in a ball of radius $r$ (given as an input parameter) identifies a ball containing $\approx t$ points. As the radius $r$ is not known in advance, we execute $O(\log |X|)$ copies of `LDP-GoodCenter` in parallel with exponentially growing values of $r$ (one of which provides a constant factor approximation to $r_{opt}$). Another alternative is to perform a binary search on $r$, hence paying in increasing round complexity to $\log \log |X|$ but saving on composition, and hence accuracy.

---

**Algorithm `LDP-GoodCenter`**

---

**Input:** Radius $r$, target number of points $t$, failure probability $\beta$, privacy parameter $\epsilon$.

**Optional input:** A predicate $\sigma : X^d \to \{0,1\}$ (otherwise set $\sigma \equiv 0$).

**Tool used:** Family $\mathcal{H}$ of $(r, c \cdot r, p, q)$-locality sensitive hash functions mapping $\mathbb{R}^d$ to a universe $U$.

**Setting:** Each player $j \in [n]$ holds a value $x_j \in X^d$. Define $S = (x_1, \ldots, x_n)$.

1. Sample a hash function $h \in \mathcal{H}$ mapping $\mathbb{R}^d$ to $U$.

2. Use Theorem 5.4 with $\frac{\epsilon}{4}$ to identify a list $L$ of length at most $32n^{1+b}/t$ containing all values $u \in U$ s.t. $|\{x \in S : h(x) = u\}| \geq \frac{t}{16} \cdot n^{-b}$.

3. Let $Z = (z_1, \ldots, z_d)$ be a random orthonormal basis of $\mathbb{R}^d$, and denote $p = 2rc\sqrt{\ln(\frac{dn}{\beta})/d}$. Partition every axis into intervals of length $p$, denoted as $\mathcal{I} = \{I_1, I_2, \ldots, I_{\sqrt{d}|X|}\}$.

4. Randomly partition $S$ into subsets $S_4^1, \ldots, S_4^d$ of size $|S_4^i| = \frac{n}{d}$. For every basis vector $z_i \in Z$, use Theorem 5.4 with $\frac{\epsilon}{4}$ to obtain for every pair $(I, u) \in \mathcal{I} \times U$ an estimation $a_i(I, u)$ for
$$|\{x \in S_4^i : \sigma(x) \neq 1 \text{ and } h(x) = u \text{ and } \langle x, z_i \rangle \in I\}|.$$

5. For every basis vector $z_i \in Z$ and for every hash value $u \in L$, denote $I(i, u) = \mathrm{argmax}_{I \in \mathcal{I}}\{a_i(I, u)\}$, and define the interval $\hat{I}(i, u)$ by extending $I(i, u)$ by $p$ to each direction (that is, $\hat{I}(i, u)$ is of length $3p$).

6. For every hash value $u \in L$, let $B(u)$ denote the box in $\mathbb{R}^d$ whose projection on every axis $z_i \in Z$ is $\hat{I}(i, u)$.

7. Randomly partition $S$ into subsets $S_7^1, \ldots, S_7^{|L|}$ of size $|S_7^\ell| = \frac{n}{|L|}$. That is, for every $u \in L$ we have a subset $S_7^u \subseteq S$. For every $u \in L$ use algorithm `LDP-AVG` with $\frac{\epsilon}{4}$ to obtain an approximation $\hat{y}_u$ for the average of $\{x \in S_7^u : h(x) = u \text{ and } x \in B(u) \text{ and } \sigma(x) \neq 1\}$.

8. Output $\{\hat{y}_u : u \in L\}$.

---

**Observation 5.14.** *Algorithm* `LDP-GoodCenter` *satisfies $\epsilon$-LDP.*

We now proceed with the utility analysis of algorithm `LDP-GoodCenter`. We will assume the existence of a family $\mathcal{H}$ of $(r, cr, p{=}n^{-b}, q{=}n^{-2-a})$-sensitive hash functions mapping $\mathbb{R}^d$ to a universe $U$, for some constants $a > b$, $r > 0$, and $c > 1$.

**Lemma 5.15.** *Fix a predicate $\sigma : X^d \to \{0, 1\}$. Let $S = (x_1, \ldots, x_n) \in (X^d)^n$ be a database which is distributed across $n$ players (each holding one row), and let $P \subseteq \{x \in S : \sigma(x) = 1\}$ be a set of $t$ input points which can be enclosed in a ball of radius $r$. Let `LDP-GoodCenter` be executed with the family $\mathcal{H}$ and with a radius $r$ and with parameters $\beta = n^{-a}/24$ and $t, \epsilon$ s.t.*

$$t \geq O\left(\frac{n^{\frac{1}{2}+b} \cdot d^{1/2}}{\epsilon} \log(dn|X|) + \frac{n^{\frac{2}{3}+b} \cdot d^{1/3}}{\epsilon^{2/3}} (\log(dn))^{2/3}\right).$$

*The algorithm outputs a set $Y$ of $\frac{32 \cdot n^{1+b}}{t}$ vectors in $\mathbb{R}^d$ such that with probability at least $n^{-a}/4$ there exists a vector $y \in Y$ such that the ball of radius $3cr$ around $y$ contains at least 1 point from $P$.*

*Proof.* First observe that, w.l.o.g., we can assume that the range $U$ of every function in $\mathcal{H}$ is of size $|U| \leq n^3$. If this is not the case, then we can simply apply a (pairwise independent) hash function with range $n^3$ onto the output of the locally sensitive hash function. Clearly, this will not decrease the probability of collusion for "close" elements (within distance $r$), and moreover, this can increase the probability of collusion for "non-close" elements (at distance at least $cr$) by at most $n^{-3} = o(n^{-2-a}) = o(q)$.

Now recall that by the properties of the family $\mathcal{H}$, for every $x, y \in \mathbb{R}^d$ s.t. $\|x - y\| \geq cr$ we have that $\Pr_{h \in \mathcal{H}}[h(x) = h(y)] \leq q = n^{-2-a}$. Using the union bound we get

$$\Pr_{h \in_R \mathcal{H}}[h(x) \neq h(y) \text{ for all } x, y \in S \text{ s.t. } \|x - y\| \geq cr] \geq (1 - n^{-a}/2).$$

Let $P \subseteq S$ denote the guaranteed set of $t$ input points $x \in S$ with $\sigma(x) \neq 1$ that are contained in a ball of radius $r$, and let $x \in P$ be an arbitrary point in $P$. By linearity of expectation, we have that

$$\mathop{\mathbb{E}}_{h \in \mathcal{H}}[|\{y \in P : h(y) \neq h(x)\}|] \leq t(1 - p) = t(1 - n^{-b}).$$

Hence, by Markov's inequality,

$$\Pr_{h \in \mathcal{H}}\left[|\{y \in P : h(y) \neq h(x)\}| \geq \frac{t(1 - n^{-b})}{1 - n^{-a}}\right] \leq 1 - n^{-a}.$$

So,

$$\Pr_{h \in \mathcal{H}}\left[|\{y \in P : h(y) = h(x)\}| \geq t\left(1 - \frac{1 - n^{-b}}{1 - n^{-a}}\right)\right] \geq n^{-a}.$$

Simplifying, for large enough $n$ (specifically, for $n^{a-b} \geq 2$) we get

$$\Pr_{h \in \mathcal{H}}\left[|\{y \in P : h(y) = h(x)\}| \geq \frac{t}{2} \cdot n^{-b}\right] \geq n^{-a}.$$

So far we have established that with probability at least $n^{-a}/2$ over the choice of $h \in \mathcal{H}$ in Step 1 the following events occur:

($E_1$) For every $x, y \in S$ s.t. $\|x - y\| \geq cr$ it holds that $h(x) \neq h(y)$; and,

($E_2$) There exists a hash value in $U$, denoted $u^*$, such that $|\{y \in P : h(y) = u^*\}| \geq \frac{t}{2} \cdot n^{-b}$.

Event ($E_1$) states that if two points in $S$ are mapped into the same hash value, then these points are close. Event ($E_2$) states that there is a "heavy" hash value $u^* \in U$, such that "many" of the points in $P$ are mapped into $u^*$. We proceed with the analysis assuming that these two events occur.

On step 2, we identify a list $L$ containing all such "heavy" hash values $u \in U$. Assuming that $t \geq O\left(\frac{1}{\epsilon} \cdot n^{0.5+b} \cdot \sqrt{\log(n/\beta) \log(1/\beta)}\right)$, Theorem 5.4 ensures that with probability at least $1 - \beta$ we have that $u^* \in L$. We continue with the analysis assuming that this is the case.

On Step 3 we generate a random orthonormal basis $Z$. By Lemma 4.2, with probability at least $(1 - \beta)$, for every $x, y \in S$ and for every $z_i \in Z$, we have that the projection of $(x - y)$ onto $z_i$ is of length at most $2\sqrt{\ln(dn/\beta)/d} \cdot \|x - y\|$. In particular, for every hash value $u \in L$ we have that the projection of $S_u \triangleq \{x \in S : h(x) = u\}$ onto every axis $z_i \in Z$ fits within an interval of length at most $p = 2rc\sqrt{\ln(dn/\beta)/d}$. Recall that we divide (on step 3) every axis $z_i \in Z$ into intervals of length $p$, denoted as $\mathcal{I} = \{I_1, I_2, \ldots\}$. Hence, for every axis $z_i \in Z$ and for every $u \in U$, we have that the projection of $S_u$ onto $z_i$ is contained within 1 or 2 consecutive intervals from $\mathcal{I}$.

On step 4 we partition $S$ into $d$ subsets $S_4^i \subseteq S$ of size $\frac{n}{d}$. By the Hoeffding bound, assuming that $t \geq 2 \cdot n^{0.5+b} \cdot \sqrt{2d \ln(\frac{2d}{\beta})}$, with probability at least $1 - \beta$, for every $i \in [d]$, we have that $|S_4^i \cap S_{u^*}| \geq \frac{|S_{u^*}|}{2d} \geq \frac{t \cdot n^{-b}}{4d}$. Recall that the projection of $S_{u^*}$ onto every axis $z_i \in Z$ fits within (at most) 2 consecutive intervals from $\mathcal{I}$. Hence, for every axis $z_i \in Z$, at least 1 interval from $\mathcal{I}$ contains at least half of the points from $S_4^i \cap S_{u^*}$, i.e., at least $\frac{t \cdot n^{-b}}{8d}$ points. Therefore, for $t \geq O\left(\frac{1}{\epsilon} \cdot n^{0.5+b} \cdot \sqrt{d \cdot \log(\frac{dn|X|}{\beta}) \log(\frac{1}{\beta})}\right)$, Theorem 5.4 ensures that with probability at least $1 - \beta$, for every $z_i \in Z$ we have that $I(i, u^*) = \operatorname{argmax}_{i \in \mathcal{I}}\{a_i(I, u^*)\}$ (defined on step 5) contains at least one point from $S_{u^*}$, and hence, the interval $\hat{I}(i, u^*)$ obtained by extending $I(i, u^*)$ by $p$ to each direction, contains (the projection of) all of the points from $S_{u^*}$ (onto the $i^{\text{th}}$ axis). As a result, the box $B(u^*)$, defined on step 7 as the box whose projection onto every axis $i$ is $\hat{I}(i, u^*)$, contains all of $S_{u^*}$. We continue with the analysis assuming that this is the case.

On step 7 we partition $S$ into $|L|$ subsets $S_7^1, \ldots, S_7^{|L|}$. For every $u \in L$ we then use $S_7^u$ to obtain a an estimation $\hat{y}_u$ for the average $y_u$ of the points $x \in S_7^u$ s.t. $h(x) = u$ and $x \in B(u)$. By the Hoeffding bound, with probability at least $1 - \beta$, we have that $|S_7^{u^*} \cap S_{u^*}| \geq \frac{|S_{u^*}|}{2|L|} \geq \frac{t^2}{128 \cdot n^{1+2b}}$. If that is the case, then as $B(u^*)$ is of side length $p$, Lemma 5.9 ensures that with probability at least $1 - \beta$ we have that $\|y_{u^*} - \hat{y}_{u^*}\|_2 \leq cr$, provided that $t \geq \frac{310}{\epsilon^{2/3}} \cdot n^{\frac{2}{3}+b} \cdot d^{1/3} \cdot \left(\log(\frac{dn}{\beta})\right)^{2/3}$. We continue with the analysis assuming that this is the case.

Observe that $y_{u^*}$ is the average of (some of) the points in $S_{u^*}$, and that every two points in $S_{u^*}$ are within distance $cr$ from each other. Hence, we get that a ball of radius $2cr$ around $y_{u^*}$ contains all of $S_{u^*}$. In particular, as $S_{u^*}$ contains at least some of the points from $P$ (the guaranteed cluster radius $r$ with $t$ input points from $S$), we have that the ball of radius $2cr$ around $y_{u^*}$ contains at least 1 point from $P$. Therefore, as $\|y_{u^*} - \hat{y}_{u^*}\|_2 \leq cr$ we get that a ball of radius $3cr$ around $\hat{y}_{u^*}$ contains at least 1 points from $P$.

Overall, with probability at least $\frac{n^{-a}}{2} - 6\beta$ we have that the output on step 8 contains at least one vector $\hat{y}$ s.t. the ball of radius $3cr$ around $\hat{y}$ contains at least one point from $P$. □

**Algorithm LDP-1Cluster**

**Input:** Radius $r$, target number of points $t$, failure probability $\beta$, privacy parameters $\epsilon, \delta$.

**Optional input:** A predicate $\sigma : X^d \to \{0, 1\}$ (otherwise set $\sigma \equiv 0$).

**Tool used:** Algorithm LDP-GoodCenter with success probability $n^{-a}/4$, and error factor (in the radius) at most $3c$.

**Setting:** Each player $j \in [n]$ holds a value $x_j \in X^d$. Define $S = (x_1, \ldots, x_n)$.

1. Denote $J = \log(\sqrt{d}|X|)$ and $K = 4 \cdot n^a \cdot \ln(1/\beta)$. Randomly partition $S$ into $R \cdot T$ subsets $S_{1,1}, \ldots, S_{R,T}$ of size $\frac{n}{JK}$ each.

2. For every $j \in [J]$ and every $k \in [K]$ apply LDP-GoodCenter on the database $S_{j,k}$ with the radius $\frac{2^j}{|X|}$, the predicate $\sigma$, privacy parameter $\frac{\epsilon}{2}$, and the parameter $t' = \frac{t}{2JK}$. Obtain a set of $L = \frac{64 \cdot n^{1+b}}{t}$ vectors: $\hat{Y}_{j,k} = \{\hat{y}_{j,k,1}, \ldots, \hat{y}_{j,k,L}\}$.

3. Randomly partition $S$ into $J \cdot K \cdot L$ subsets $S_{1,1,1}, \ldots, S_{J,K,L}$ of size $\frac{n}{JKL}$ each.

4. For every $j \in [J]$, every $k \in [K]$, and every $\ell \in [L]$, use Theorem 5.3 on the database $S_{j,k,\ell}$ with $\frac{\epsilon}{2}$ to obtain an estimation $\hat{c}_{j,k,\ell}$ for the number of points $x \in S_{j,k,\ell}$ enclosed in a ball of radius $5c\frac{2^j}{|X|}$ around $\hat{y}_{j,k,\ell}$, for which $\sigma(x) \neq 1$.

5. Let $j'$ be the smallest number in $[J]$ for which there exist $(k', \ell') \in [K] \times [L]$ s.t.

$$JKL \cdot \hat{c}_{j',k',\ell'} \geq t - \frac{224}{\epsilon} \cdot \frac{n^{1+a}}{t^{1/2}} \cdot \sqrt{\log(d|X|) \log(\frac{1}{\beta}) \log(\frac{32n}{\beta})}.$$

Return the vector $\hat{t}_{j',k',\ell'}$ and the radius $5c\frac{2^{j'}}{|X|}$.

We are now ready to prove Theorem 5.13. The proof is via the construction of algorithm `LDP-1Cluster`.

*Proof of Theorem 5.13.* The privacy properties of algorithm `LDP-1Cluster` are immediate. We now proceed with the utility analysis.

Let $r_{opt}$ be the radius of the smallest ball in $X^d$ containing at least $t$ points from $\{x \in S : \sigma(x) \neq 1\}$, and let $P$ denote a set of $t$ such points. Let $j^*$ denote the smallest integer s.t. $r_{opt} \leq \frac{2^{j^*}}{|X|}$, and denote $r = \frac{2^{j^*}}{|X|}$.

On step 1 we partition $S$ into $JK$ subsets $S_{1,1}, \ldots, S_{J,K}$. By the Hoeffding bound, assuming that $t \geq 8 \cdot n^{\frac{1+a}{2}} \cdot \sqrt{\log(d|X|) \log(\frac{1}{\beta}) \log(\frac{8n}{\beta})}$, with probability at least $1 - \beta$, for every $k \in [K]$ we have that $|S_{j^*,k} \cap P| \geq t' = \frac{t}{2JK}$. That is, for every $k \in [K]$, the database $S_{j^*,k}$ contains at least $t'$ points from $P$, which can be enclosed in a ball of radius $\frac{2^{j^*}}{|X|} = r$. Hence, by the guarantees of algorithm `LDP-GoodCenter` (Theorem 5.15), for every $k \in [K]$, with probability at least $n^{-a}/4$ we have that $\hat{Y}_{j^*,k}$ contains a vector $\hat{y}$ s.t. the ball of radius $3cr$ around $\hat{y}$ contains at least 1 point from $P$, provided that

$$t \geq O\left(\frac{n^{\frac{1}{2}+a+b} \cdot d^{1/2}}{\epsilon} \log(dn|X|) \sqrt{\log(\frac{1}{\beta})} + \frac{n^{\frac{2}{3}+a+b} \cdot d^{1/3}}{\epsilon^{2/3}} \log(dn|X|) \left(\log(\frac{1}{\beta})\right)^{1/3}\right).$$

Therefore, as $K = 4n^a \ln(\frac{1}{\beta})$, we get that with probability at least $(1 - \beta)$, the set $\bigcup_{k \in [K]} \hat{Y}_{j^*,k}$ contains at least one vector, denoted $\hat{y}_{j^*,k^*,\ell^*}$, s.t. the ball of radius $3cr$ around it contains at least 1 point from $P$. We continue with the analysis assuming that this is the case. Observe that, as the set $P$ is of diameter (at most) $2r$, we get that a ball of radius $5cr$ around $\hat{y}_{j^*,k^*,\ell^*}$ contains *all* of $P$.

On step 3 we partition $S$ into $JKL$ subsets $S_{1,1,1}, \ldots, S_{J,K,L}$. For every $(j, k, \ell) \in [J] \times [K] \times [L]$, we then use Theorem 5.3 to obtain an estimation $\hat{c}_{j,k,\ell}$ for the number of points from $S_{j,k,\ell}$ that are enclosed in the ball of radius $5c\frac{2^j}{|X|}$ around $\hat{y}_{j,k,\ell}$. Let us denote the true number of points from $S_{j,k,\ell}$ in that ball as $c_{j,k,\ell}(S_{j,k,\ell})$. Similarly, let $c_{j,k,\ell}(S)$ denote the number of points from $S$ in that ball, i.e., the number of points from $S$ that are enclosed in the ball of radius $5c\frac{2^j}{|X|}$ around $\hat{y}_{j,k,\ell}$.

By the Hoeffding bound, with probability at least $1 - \beta$, for every $(j, k, \ell) \in [J] \times [K] \times [L]$ we have that $|c_{j,k,\ell}(S) - JKL \cdot c_{j,k,\ell}(S_{j,k,\ell})| \leq 32 \cdot \frac{n^{1+a}}{t^{1/2}} \cdot \sqrt{\log(d|X|) \log(\frac{1}{\beta}) \log(\frac{32n}{\beta})}$. In addition, by the guarantees of Theorem 5.3, with probability at least $1 - \beta$, for every $(j, k, \ell) \in [J] \times [K] \times [L]$ we have that $|JKL \cdot c_{j,k,\ell}(S_{j,k,\ell}) - JKL \cdot \hat{c}_{j,k,\ell}| \leq \frac{192}{\epsilon} \cdot \frac{n^{1+a}}{t^{1/2}} \cdot \sqrt{\log(d|X|) \log(\frac{1}{\beta}) \log(\frac{7n}{\beta})}$, and hence, for every $(j, k, \ell) \in [J] \times [K] \times [L]$ we have

$$|c_{j,k,\ell}(S) - JKL \cdot \hat{c}_{j,k,\ell}| \leq \frac{224}{\epsilon} \cdot \frac{n^{1+a}}{t^{1/2}} \cdot \sqrt{\log(d|X|) \log(\frac{1}{\beta}) \log(\frac{32n}{\beta})}.$$

We continue with the analysis assuming that this is the case.

We have already established the existence of a vector $\hat{y}_{j^*,k^*,\ell^*} \in \bigcup_{k \in [K]} \hat{Y}_{j^*,k}$ s.t. the ball of radius $5c\frac{2^{j^*}}{|X|}$ around it contains all of $P$, i.e., contains at least $t$ points from $S$. That is, $c_{j^*,k^*,\ell^*}(S) \geq t$, and so

$$JKL \cdot \hat{c}_{j^*,k^*,\ell^*} \geq t - \frac{224}{\epsilon} \cdot \frac{n^{1+a}}{t^{1/2}} \cdot \sqrt{\log(d|X|) \log(\frac{1}{\beta}) \log(\frac{32n}{\beta})}. \quad (4)$$

Hence, as $j', k', \ell'$ is defined on step 5 as the minimal $j$ for which Inequality (4) holds, we get that $j' \leq j^*$, and hence, $\frac{2^{j'}}{|X|} \leq 2r_{opt}$. That is, the radius returned at the end of the execution is at most $10c \cdot r_{opt}$. In addition, as $j'$ (from step 5) satisfies Inequality (4), we get that the number of points from $S$ contained in the ball of radius $10c \cdot r_{opt}$ around $\hat{y}_{j',k',\ell'}$ is at least

$$t - \frac{448}{\epsilon} \cdot \frac{n^{1+a}}{t^{1/2}} \cdot \sqrt{\log(d|X|)\log(\frac{1}{\beta})\log(\frac{32n}{\beta})}.$$

$\square$

## 6 Application to $k$-means clustering

For a set of centers $C \subseteq \mathbb{R}^d$ and a point $x \in \mathbb{R}^d$ we denote $\text{dist}(x, C) = \min_{c \in C} \|x - c\|_2$. For a finite set $S \subseteq \mathbb{R}^d$, the sum of squared distances is defined as

$$\text{cost}(S, C) = \sum_{x \in S} \text{dist}^2(x, C).$$

If $C = \{c\}$ we denote $\text{cost}(S, c) = \text{cost}(S, \{c\})$ for simplicity. For a *weighted* set $S = \{(x_1, \alpha_1), \ldots, (x_\ell, \alpha_\ell)\}$, the weighted cost is

$$\text{cost}(S, C) = \sum_{(x,\alpha) \in S} \alpha \cdot \text{dist}^2(x, C).$$

**Definition 6.1** ($k$-means). *Let $S$ be a (weighted or unweighted) finite set of points in $\mathbb{R}^d$. A set $C^*$ of $k$ centers (points) in $\mathbb{R}^d$ is called $k$-means of $S$ if it minimizes $\text{cost}(S, C)$ over every such set $C$.*

**Definition 6.2** (Approximated $k$-means). *Let $S$ be a (weighted or unweighted) finite set of points in $\mathbb{R}^d$. A set $C$ of $k$ centers in $\mathbb{R}^d$ is a $(\gamma, \eta)$-approximation for the $k$-means of $S$ if*

$$\text{cost}(S, C) \leq \gamma \cdot \text{cost}(S, C^*) + \eta,$$

*where $C^*$ is a $k$-means of $S$. When $\eta = 0$ we omit it and say that $C$ is a $\gamma$-approximation for the $k$-means of $S$.*

### 6.1 $k$-means clustering – the centralized setting

As we mentioned in the introduction, Feldman et al. [15] showed that private algorithms for the 1-cluster problem translate to private algorithm for approximating the $k$-means of the data. Specifically, they showed the following theorem:

**Theorem 6.3** ([15]). *Fix a domain $X^d$ and a number $k \geq 1$. Assume the existence of an $(\epsilon, \delta)$-differentially private algorithm that, for every $n \geq n_{min}$, solves the 1-cluster problem $(X^d, n, t = \frac{3n}{8k})$ with parameters $(\Delta, w)$ and error probability $\beta$, where $\Delta \leq \frac{n}{8k}$. Then, for every $\delta' > 0$, there exists an $(\epsilon', 4k\log(n)\delta + \delta')$-differentially private algorithm such that the following holds. Given a database $S \in (X^d)^n$, with probability at least $1 - O(\beta \cdot k \cdot \log(n))$, the algorithm returns a $(\gamma, \eta)$-approximation to the $k$-means of $S$, where*

$$\gamma = O(w^2 \cdot k), \qquad \eta = O(n_{min}) \cdot \text{diam}(X^d), \qquad \epsilon' = \sqrt{8k \log(n) \log(\frac{1}{\delta'})}\epsilon + 4k \log(n)\epsilon(e^\epsilon - 1).$$

Using this transformation with the (private) algorithm of [21] for the 1-cluster problem, Feldman et al. [15] obtained the following result.[3]

**Theorem 6.4** ([15]). *Fix $\beta, \epsilon, \delta \leq 1$. There exists an $(\epsilon, \delta)$-differentially private algorithm that, given a database $S \in (X^d)^n$, identifies with probability $1 - \beta$ a $(\gamma, \eta)$-approximation for the k-means of $S$, where $\gamma = O(k \log(n))$ and $\eta = \tilde{O}\left(\frac{1}{\epsilon} \cdot d^{0.5} \cdot k^{1.5} \cdot \mathrm{diam}(X^d)\right)$.*

Applying the transformation of Theorem 6.3 with our new algorithm for the 1-cluster problem, we obtain the following result.

**Theorem 6.5.** *Fix $\beta, \epsilon, \delta \leq 1$. There exists an $(\epsilon, \delta)$-differentially private algorithm that, given a database $S \in (X^d)^n$, identifies with probability $1 - \beta$ a $(\gamma, \eta)$-approximation for the k-means of $S$, where $\gamma = O(k)$ and $\eta = \tilde{O}\left(\frac{1}{\epsilon^{1.01}} \cdot d^{0.51} \cdot k^{1.51} \cdot \mathrm{diam}(X^d)\right)$.*

## 6.2 $k$-means clustering – the distributed setting

Towards obtaining our LDP algorithm for $k$-means, we start by presenting a non-private algorithm that compute an approximation for the $k$-means (in fact, this is a *family* of algorithms, as the algorithm can make arbitrary decisions throughout the execution). Afterwards we will show that (a successful run of) our LDP algorithm can be identified with a possible execution of the non-private algorithm, showing that our private algorithm indeed computes an approximation for the $k$-means. The non-private algorithm we use was first introduced by [13], and was modified by [15] to obtain the transformation from 1-cluster to $k$-means in the centralized model.

**Theorem 6.6** ([13, 15]). *Every execution of algorithm* `Non-Private-k-Mean` *on a database $S \in (X^d)^n$ with parameters $\nu, w, t_{min}$, returns a set $C$ of $k$ centers such that*

$$\mathrm{cost}(S, C) \leq O(w^2 k) \cdot \mathrm{cost}(S, C^*) + O\left(\nu k \log(n) + t_{min}\right) \cdot \mathrm{diam}(X^d),$$

*where $C^*$ are the optimal k-means of the input database $S$.*

The analysis of algorithm `Non-Private-k-Mean` is almost identical to the analysis of [13, 15]. The only modification is that we allowed some "slackness" in the choices for $G_i$ and $\alpha_i$, which will be helpful for our LDP variant of the algorithm. The proof of Theorem 6.6 appears in the appendix for completeness.

**Lemma 6.7.** `LDP-k-Mean` *satisfies $(\epsilon', \delta)$-LDP for $\epsilon' = \sqrt{24k \log(n) \ln(1/\delta)}\epsilon + 12k \log(n)\epsilon(e^\epsilon - 1)$.*

Lemma 6.7 follows directly from Theorem 2.9 (advanced composition). We now present the utility analysis.

**Lemma 6.8.** *Let $S = (x_1, \ldots, x_n) \in (X^d)^n$ be a database which is distributed across $n$ players (each holding one point in $X^d$), and denote $\nu = \frac{16}{\epsilon}\sqrt{n \cdot \log(8/\beta)}$.*

*Let algorithm* `LDP-k-Mean` *be executed with an algorithm for the 1-cluster problem with parameters $w, \Delta, t_{min}$, such that $t_{min} \geq 9\nu + 6\Delta$. Except with probability at most $O(\beta \cdot k \log n)$, the output $C$ is s.t. there exists an execution of algorithm* `Non-Private-k-Mean` *on the database $S$ with parameters $\nu$ and $w$ and $t'_{min} \leq \frac{8k}{3} \cdot t_{min} + \nu$ that results in the same output $C$.*

---

[3]For simplicity, the $\tilde{O}$ notation in theorems 6.4 and 6.5 hides factors of $\log(\frac{kn}{\beta\delta})$ and $2^{O(\log^* |X|)}$ from the statement on the additive error $\eta$.

**Algorithm Non-Private-k-Mean** [13, 15]

**Inputs:** A set of $n$ input points $S \in (X^d)^n$ and an integer $k \geq 1$.

**Parameter:** $\nu, w, t_{min}$.

**Output:** A set $C$ of $k$ centers.

1. $B \leftarrow \emptyset$.

2. While $|S| > t_{min}$ do (let $i$ denote the index of the current iteration)

    (a) $n_i \leftarrow |S|$.

    (b) Let $r_{i,opt}$ denote the radius of a smallest ball that contains $\frac{3n_i}{8k}$ points from $S$.

    (c) Let $c_i \in X^d$ and $G_i \subseteq S$ be s.t.

       i. $G_i$ is contained within a ball of radius $r_i \leq w \cdot r_{i,opt}$ around $c_i$,
       ii. $\frac{n_i}{4k} \leq |G_i| \leq \frac{3n_i}{8k}$.

    (d) $S \leftarrow S \setminus G_i$.

    (e) Arbitrarily choose $|G_i| - \nu \leq \alpha_i \leq |G_i|$.

    (f) $B \leftarrow B \cup \{(c_i, \alpha_i)\}$.

3. $C \leftarrow$ an $O(1)$-approximation to the $k$-means of $B$.

4. Return $C$.

---

Combining Lemmas 6.7 and 6.8 with our LDP algorithm for the 1-cluster problem, yields the following theorem.[4]

**Theorem 6.9.** *Fix $\beta, \epsilon, \delta \leq 1$. There exists an $(\epsilon, \delta)$-LDP algorithm that, for a (distributed) database $S \in (X^d)^n$, identifies with probability $1 - \beta$ a $(\gamma, \eta)$-approximation for the $k$-means of $S$, where $\gamma = O(k)$ and $\eta = \tilde{O}\left(\frac{1}{\epsilon} \cdot n^{0.67} \cdot d^{1/3} \cdot k^{0.5} \cdot \operatorname{diam}(X^d)\right)$.*

*Proof of Lemma 6.8.* We say that an application of step 4a is *successful* if it identifies a center $c_i$ and a radius $r_i$ s.t. the ball of radius $r_i$ around $c_i$ contains at least $(t_i - \Delta)$ points from $S_i$, and furthermore, $r_i \leq w \cdot \hat{r}_{i,opt}$ (where $\hat{r}_{i,opt}$ is the radius of a smallest ball enclosing $t_i$ points from $S_i$). Similarly, we say that the applications of steps 4b,4c,4e,4f are *successful* if they result in estimations satisfying the stated requirements. As every such application succeeds with probability $(1 - \beta)$, and as there are at most $4k \log n$ iterations, we have that all of these applications succeed together with probability at least $1 - O(\beta \cdot k \log n)$. We proceed with the analysis assuming that this is true, in which case we say that the entire execution is *successful*.

The proof proceeds by showing that for every successful execution of algorithm LDP-k-Mean there exists an execution of algorithm Non-Private-k-Mean that results in the same output $C$. Hence, successful executions of algorithm LDP-k-Mean enjoy the utility guarantees of algorithm Non-Private-k-Mean.

Fix a successful execution of algorithm LDP-k-Mean, denoted as $Exec_1$ (this amounts to fixing all of the randomness throughout the computation). We analyze the set of points that are excluded

---

[4]For simplicity, the $\tilde{O}$ notation in Theorem 6.9 hides factors of $\log(\frac{kn|X|}{\beta\delta})$ from the additive error $\eta$.

**Algorithm LDP-k-Mean**

**Inputs:** Privacy parameters $\epsilon, \delta$, failure probability $\beta$, and desired number of centers $k$.

**Oracle:** LR Oracle access to a database $S \in (X^d)^n$.

**Tool used:** An $\epsilon$-LDPns algorithm for the 1-cluster problem with parameters $w, \Delta, t_{min}$.

**Output:** A set $C$ of $k$ centers.

1. $B \leftarrow \emptyset$.

2. Define $\sigma : X^d \rightarrow \{0, 1\}$ where $\sigma(x) \equiv 0$. We represent $\sigma$ as a subset of $X^d$, i.e., $\sigma = \emptyset$.

3. Set $i = 1$. Denote $n_1 = n$ and $\hat{n}_1 = n$. Also denote $\nu = \frac{16}{\epsilon}\sqrt{n_1 \cdot \log(8/\beta)}$.
   
   % We assume that $t_{min} \geq 9\nu + 6\Delta$. Otherwise set $t_{min} = 9\nu + 6\Delta$.

4. While $(\hat{n}_i > \frac{8k}{3}t_{min})$ and $(i \leq 4k \log n)$ do

   (a) Use the algorithm for the 1-cluster problem with the predicate $\sigma$ and parameter $t_i = \frac{3\hat{n}_i}{8k}$ to obtain a center $c_i$ and a radius $r_i$.
   
   % Let $S_i$ denote the data of all users holding an input element $x$ s.t. $\sigma(x) \neq 1$. With probability $(1 - \beta)$ we have that the ball of radius $r_i$ around $c_i$ contains at least $(t_i - \Delta)$ points from $S_i$. Furthermore, $r_i \leq w \cdot \hat{r}_{i,opt}$, where $\hat{r}_{i,opt}$ is the radius of a smallest ball enclosing $t_i$ points from $S_i$.

   (b) Let $b_i$ denote the number of users holding an input element $x$ s.t. $\sigma(x) \neq 1$ and $\|x - c_i\|_2 \leq r_i$. Let $\hat{b}_i$ be a noisy estimation (satisfying $\epsilon$-LDP, e.g., using Theorem 5.3) for $b_i$, such that w.p. $(1 - \beta)$ we have that $b_i - \nu \leq \hat{b}_i \leq b_i$.

   (c) Let $h_i : X^d \rightarrow \{0, 1\}$ be a random hash function s.t. for every $x \in X^d$ we have that $\Pr[h_i(x) = 1] = \min\left\{1, \frac{5\hat{n}_i}{16k\hat{b}_i}\right\} \triangleq \hat{p}_i$.
   
   % It suffices to use a $\lambda$-wise independent hash function for $\lambda = \ln(1/\beta)$. Let $\widetilde{G}_i$ denote the set of input points $x \in S_i$ such that $\|x - c_i\|_2 \leq r_i$ and $h_i(x) = 1$. Using a standard tail bound for sum of $\lambda$-wise independent random variables (see, e.g., Lemma 5.6), with probability $(1 - \beta)$ we have that $|\widetilde{G}_i| \in \hat{p}_i \cdot b_i \pm \sqrt{8b_i \ln(1/\beta)}$.

   (d) Let $\sigma \leftarrow \sigma \cup \{x \in X^d : \|x - c_i\|_2 \leq r_i$ and $h_i(x) = 1\}$.
   
   % That is, roughly a $\hat{p}_i$ fraction of the points in the found ball are excluded from the rest of the computation.

   (e) $B \leftarrow B \cup \{(c_i, \hat{\alpha}_i)\}$, where $\hat{\alpha}_i = \hat{b}_i \cdot \hat{p}_i - \nu$.
   
   % Let $\alpha_i$ denote the number of users that were excluded on the last step. With high probability we will have that $\alpha_i - 3\nu \leq \hat{\alpha}_i \leq \alpha_i$

   (f) Set $i \leftarrow (i + 1)$, and let $n_i$ denote the number of users that are not excluded from the computation, i.e., $n_i = |\{x \in S : \sigma(x) \neq 1\}|$. Let $\hat{n}_i$ be a noisy estimation (satisfying $\epsilon$-LDP, e.g., using Theorem 5.3) for $n_i$, such that w.p. $(1-\beta)$ we have that $n_i - \nu \leq \hat{n}_i \leq n_i$.

5. $C \leftarrow$ an $O(1)$-approximation to the $k$-means of $B$.

6. Return $C$.

on step 4d of iteration $i$ of $Exec_1$, denoted as $\widetilde{G}_i$ (that is, $\widetilde{G}_i$ is the set of input points $x \in S_i$ such that $\|x - c_i\|_2 \leq r_i$ and $h_i(x) = 1$). We next show that, for every $i$, the set $\widetilde{G}_i$ satisfies conditions (i),(ii) of step 2c of algorithm Non-Private-k-Mean. For condition (i), recall that as step 4a succeeded, the set $\widetilde{G}_i$ is contained within a ball of radius $w \cdot \hat{r}_{i,opt}$ around $c_i$, where $\hat{r}_{i,opt}$ is the radius of a smallest ball enclosing $t_i = \frac{3\hat{n}_i}{8k}$ points from $S_i$. Let $r_{i,opt}$ denote the radius of a smallest ball enclosing $\frac{3n_i}{8k}$ points from $S_i$. As $\hat{n}_i \leq n_i$ we have that $\hat{r}_{i,opt} \leq r_{i,opt}$, and hence, $\widetilde{G}_i$ is contained within a ball of radius $w \cdot r_{i,opt}$ around $c_i$, satisfying condition (i).

Let $b_i$ and $\hat{b}_i$ be as in step 4b of the $i^{\text{th}}$ iteration of $Exec_1$ (that is, $b_i$ denotes the number of users holding an input element $x$ s.t. $\sigma(x) \neq 1$ and $\|x - c_i\|_2 \leq r_i$, and $\hat{b}_i$ is a noisy estimation satisfying $b_i - \nu \leq \hat{b}_i \leq b_i$). As the execution is successful, we have that

$$b_i \geq \frac{3\hat{n}_i}{8k} - \Delta \geq \frac{3n_i}{8k} - \frac{3\nu}{8k} - \Delta. \tag{5}$$

Assuming that $t_{min} \geq 9\nu + 6\Delta$, we have that

$$n_i \geq \hat{n}_i \geq \frac{8k}{3} t_{min} \geq 24k\nu + 16k\Delta. \tag{6}$$

Plugging this into Inequality (5) we get that

$$b_i \geq \frac{5n_i}{16k} + \nu. \tag{7}$$

Recall that on step 4c we define $\hat{p}_i \triangleq \min\left\{1, \frac{5\hat{n}_i}{16k\hat{b}_i}\right\}$. By Inequality (7) and by the fact that $n_i - \nu \leq \hat{n}_i \leq n_i$ and $b_i - \nu \leq \hat{b}_i \leq b_i$, we get that

$$\frac{5n_i}{16kb_i} - \frac{\nu}{b_i} \leq \hat{p}_i \leq \frac{5n_i}{16kb_i} + \frac{\nu}{b_i}. \tag{8}$$

Finally, as step 4c succeeded, we have that

$$\hat{p}_i \cdot b_i - \sqrt{8b_i \ln(1/\beta)} \leq |\widetilde{G}_i| \leq \hat{p}_i \cdot b_i + \sqrt{8b_i \ln(1/\beta)}, \tag{9}$$

which in combination with Inequalities (8) and (6) yields:

$$\frac{n_i}{4k} \leq |\widetilde{G}_i| \leq \frac{3n_i}{8k}. \tag{10}$$

So, every iteration of $Exec_1$ identifies a set $\widetilde{G}_i$ satisfying conditions (i) and (ii) of algorithm Non-Private-k-Mean, and excludes it from future iterations. In addition, by the fact that $b_i - \nu \leq \hat{b}_i \leq b_i$ and by Inequality (9), we get that the weight $\hat{\alpha}_i$ (from step 4e) satisfies $|\widetilde{G}_i| - 3\nu \leq \hat{\alpha}_i \leq |\widetilde{G}_i|$, as is required on step 2e of algorithm Non-Private-k-Mean. This shows that $Exec_1$ has a matching execution of algorithm Non-Private-k-Mean such that the set of centers $B$ is identical throughout both of the executions. We next show that the stopping condition of both executions can also be unified.

We have already established that in every iteration $i$ of $Exec_1$ we have that $|\widetilde{G}_i| \geq \frac{n_i}{4k}$. That is, in every iteration, at least $\frac{n_i}{4k}$ points are excluded from $S$, and hence, the number of iterations is at most $4k \log n$. Thus, the loop of step 4 only halts when $\hat{n}_i \leq \frac{8k}{3} \cdot t_{min}$ (and not because of the condition on the number of iterations). As we next explain, this can be identified with the stopping condition in `Non-Private-k-Mean` for a slightly different value $t'_{min}$.

Consider the iteration $i$ in which `LDP-k-Mean` halts. In that iteration we had that $\hat{n}_i \leq \frac{8k}{3} \cdot t_{min}$ and that $\hat{n}_i \geq n_i - \nu$. Hence, $n_i \leq \frac{8k}{3} \cdot t_{min} + \nu$. In addition, as $n_i < n_{i-1}$, there exists a number $t'_{min}$ such that $n_i \leq t'_{min} < n_{i-1}$ and $t'_{min} \leq \frac{8k}{3} \cdot t_{min} + \nu$. The execution $Exec_1$ would remain intact had we replaced the condition of the while loop with $(n_i \geq t'_{min})$.

This shows that $Exec_1$ has a matching execution of algorithm `Non-Private-k-Mean`, denoted $Exec_2$, such that both executions have exactly the same number of iterations, and furthermore, the set of centers $B$ is identical in $Exec_1$ and in $Exec_2$ on every step of the executions. In particular, this is the case at the end of the executions, which results in the same output $C$. □

# A Missing proofs

**Lemma A.1.** *Algorithm* `LSH-GoodCenter` *preserves* $(2\epsilon, 2\delta)$*-differential privacy.*

*Proof.* Algorithm LSH-GoodCenter interacts with its input database on steps 2, 4a, 5, 6, and 7. Step 2 invokes the algorithm from Theorem 2.6 which is $(\frac{\epsilon}{4}, \frac{\delta}{4})$-differentially private, to obtain $L \subset U$. For every $u \in L$ we make (in Step 4a) $d$ applications of the algorithm from Theorem 2.6 that interact only with $S_u$ (and no other points in $S$). By the advanced composition (Theorem 2.9) this preserves $(\frac{\epsilon}{4}, \frac{\delta}{4})$-differential privacy w.r.t. $S_u$. As the sets $S_u$ and $S_{u'}$ are mutually disjoint, Step 4a preserves $(\frac{\epsilon}{2}, \frac{\delta}{2})$-differential privacy overall. Similarly, Step 5 interacts with the data for at most $n$ times (applying Gaussian mechanism). As each application of the Gaussian mechanism is $(\frac{\epsilon}{4}, \frac{\delta}{4})$-differentially private, we get that Step 5 preserves $(\frac{\epsilon}{2}, \frac{\delta}{2})$-differential privacy. Steps 6, and 7 initialize and use Algorithm `AboveThreshold`, which is $(\frac{\epsilon}{4}, 0)$-differentially private. Overall, `GoodCenter` is $(2\epsilon, 2\delta)$-differentially private by application of simple composition (Theorem 2.8). □

**Lemma A.2.** *Suppose $X_1, \ldots, X_n$ are independent samples from* $\text{Lap}(\frac{1}{\epsilon})$. *That is, every $X_i$ has probability density function $f(x) = \frac{\epsilon}{2} \exp(-\epsilon |x|)$. Let $X = X_1 + \cdots + X_n$. For every $t \geq 0$ it holds that*

$$\Pr[|X| \geq t] \leq 6 \exp\left(-\frac{\epsilon \cdot t}{\sqrt{2n}}\right).$$

*Moreover, for every $0 \leq t < 2n/\epsilon$ we have that*

$$\Pr[|X| \geq t] \leq 2 \exp\left(-\frac{\epsilon^2 \cdot t^2}{4n}\right).$$

The proof of Lemma A.2 is standard. We include a proof here for completeness as we are unable to find an appropriate reference.

*Proof of Lemma A.2.* By symmetry, it suffices to analyze $\Pr[X \geq t]$. Indeed, for every $c > 0$ we have that

$$\begin{aligned}
\Pr[X \geq t] &= \Pr[cX \geq ct] = \Pr\left[e^{cX} \geq e^{ct}\right] \\
&\leq e^{-ct} \cdot \mathbb{E}\left[e^{cX}\right] = e^{-ct} \cdot \left(\mathbb{E}\left[e^{cX_1}\right]\right)^n.
\end{aligned} \qquad (11)$$

Furthermore,

$$\begin{aligned}
\mathbb{E}\left[e^{cX_1}\right] &= \int_0^\infty \Pr\left[e^{cX_1} \geq z\right] dz = \int_0^\infty \Pr\left[X_1 \geq \frac{1}{c} \ln z\right] dz \\
&= \int_0^1 \Pr\left[X_1 \geq \frac{1}{c} \ln z\right] dz + \int_1^\infty \Pr\left[X_1 \geq \frac{1}{c} \ln z\right] dz \\
&= \int_0^1 \left[1 - \frac{1}{2} \exp\left(\frac{\epsilon \cdot \ln z}{c}\right)\right] dz + \int_1^\infty \left[\frac{1}{2} \exp\left(-\frac{\epsilon \cdot \ln z}{c}\right)\right] dz \\
&= 1 - \frac{1}{2} \int_0^1 z^{\epsilon/c} dz + \frac{1}{2} \int_1^\infty z^{-\epsilon/c} dz = \frac{1}{1 - (c/\epsilon)^2},
\end{aligned}$$

where the last equality follows by asserting that $\epsilon/c > 1$ and solving the integrals. Plugging into Equality (11) we get that

$$\Pr[X \geq t] \leq e^{-ct} \cdot \left(\mathbb{E}\left[e^{cX_1}\right]\right)^n = e^{-ct} \cdot (1 - (c/\epsilon)^2)^{-n} \leq e^{-ct} \cdot e^{2n \cdot (\frac{c}{\epsilon})^2},$$

where the last inequality holds whenever $\epsilon/c \geq 1.2$. The lemma now follows by choosing $c = \frac{\epsilon^2 t}{4n}$ (in which case $\epsilon/c \geq 2$ for every $t \leq 2n/\epsilon$) or by choosing $c = \epsilon/\sqrt{2n}$. □

# B Proof of Theorem 6.6 [13, 15]

| | |
|---|---|
| $C^*$ | A set of $k$ centers that minimizes $\text{cost}(S, C)$ over the original database $S$ over every such $C$. |
| $S_i$ | The database $S$ during the $i^{\text{th}}$ iteration. |
| $S_i^*$ | The $3|S_i|/4$ points $x \in S_i$ with smallest value $\text{dist}(x, C^*)$. |

Table 2: Notations for the analysis of algorithm `Non-Private-k-Mean`

**Lemma B.1.** *In every iteration $i$ we have that* $\text{cost}(G_i, c_i) \leq w^2 \cdot \text{cost}(S_i^*, C^*)$

*Proof.* By the pigeonhole principle there must be a center $x \in C^*$ that serves $m \geq \frac{|S_i^*|}{k}$ points in $S_i^*$, i.e., at least $1/k$ fraction of the points in $S_i^*$ have $x$ as their closest center in $C^*$. At least half of them (i.e., $\frac{|S_i^*|}{2k}$) have distance at most $\tilde{r} = \sqrt{\frac{2\text{cost}(S_i^*, C^*)}{m}}$ to $x$, as otherwise we would get that $\text{cost}(S_i^*, C^*) > \frac{|S_i^*|}{2k} \cdot \frac{2\text{cost}(S_i^*, C^*)}{m} \geq \text{cost}(S_i^*, C^*)$.

We conclude that there is a ball of radius $\tilde{r}$ that contains at least $m/2 \geq \frac{|S_i^*|}{2k} = \frac{3|S_i|}{8k}$ points from $S_i$. By definition of $r_{i,opt}$, we have that

$$r_{i,opt} \leq \tilde{r} = \sqrt{\frac{2\text{cost}(S_i^*, C^*)}{m}}.$$

That is,

$$\frac{m}{2} r_{i,opt}^2 \leq \text{cost}(S_i^*, C^*).$$

Hence,

$$\text{cost}(G_i, c_i) \leq |G_i| \cdot (w \cdot r_{i,opt})^2 \leq \frac{3|S_i|}{8k} \cdot (w \cdot r_{i,opt})^2 = \frac{|S_i^*|}{2k} \cdot (w \cdot r_{i,opt})^2 \leq w^2 \cdot \frac{m}{2} r_{i,opt}^2 \leq w^2 \cdot \text{cost}(S_i^*, C^*),$$

where the first and second inequalities follow from the conditions on $G_i$ on step 2(c) of the algorithm. □

**Lemma B.2.**
$$\sum_{i=1}^{|B|} \text{cost}(G_i, c_i) \leq O(w^2 \cdot k) \cdot \text{cost}(S, C^*).$$

*Proof.* We order the points in $S$ by $S = (x_1, x_2, \ldots, x_n)$, such that $\text{dist}(x_{j_1}, C^*) \leq \text{dist}(x_{j_2}, C^*)$ for every $j_1 < j_2$, where ties are broken arbitrarily. Let

$$\begin{aligned} U_i &= \{x_1, x_2, \ldots, x_{n-|S_i|}\}, \\ V_i &= \{x_{n-|S_i|+1}, \ldots, x_{n-|S_i|+|S_i^*|}\}. \end{aligned}$$

Clearly, $|U_i \cup V_i| = |U_i| + |V_i|$ (as the sets $U_i$ and $V_i$ are disjoint). Moreover, as there are at most $(n - |S_i|)$ elements outside of $S_i$, we have that

$$\begin{aligned}|(U_i \cup V_i) \cap S_i| &\geq |U_i| + |V_i| - (n - |S_i|)\\ &= |V_i| = |S_i^*|.\end{aligned}$$

So, the sets $(U_i \cup V_i)$ and $S_i$ have at least $|S_i^*|$ elements in common. As $S_i^*$ contains the values $x \in S_i$ with smallest distance $\text{dist}(x, C^*)$, and as $(U_i \cup V_i)$ contain all such elements from $S$, we get that these common elements contain $S_i^*$, that is, $S_i^* \subseteq (U_i \cup V_i)$.

The set $V_i$ contains the $|S_i^*|$ points $x \in (U_i \cup V_i)$ with the largest values $\text{dist}(x, C^*)$. Hence, $\text{cost}(S_i^*, C^*) \leq \text{cost}(V_i, C^*)$. Combining this with Lemma B.1, and summing over every $i$ we get

$$\sum_{i=1}^{|B|} \text{cost}(G_i, c_i) \leq \sum_{i=1}^{|B|} w^2 \cdot \text{cost}(V_i, C^*).$$

We now want to relate the right hand side of the above inequality to the input points $S$. Specifically, we want to show that every point $x \in S$ is included in at most $O(k)$ sets $V_i$. To that end, observe that:

1. The index of the last point in $V_i$ is $n - |S_i| + |S_i^*| = n - \frac{|S_i|}{4}$.

2. The index of the first point in $V_{i+12k}$ is $n - |S_{i+12k}| + 1 \geq n - |S_i| \cdot (1 - \frac{1}{4k})^{6k} > n - \frac{|S_i|}{4}$.

Hence, every point appears in $O(k)$ sets in the sequence $V_1, V_2, \ldots$. Hence,

$$\sum_{i=1}^{|B|} \text{cost}(G_i, c_i) \leq \sum_{i=1}^{|B|} w^2 \cdot \text{cost}(V_i, C^*) \leq O(w^2 \cdot k) \cdot \text{cost}(S, C^*).$$

□

**Lemma B.3.**

$$\text{cost}(S, C) \leq O(w^2 k) \cdot \text{cost}(S, C^*) + O\left(t_{min} \cdot \sqrt{d} + \nu k \log(n) \sqrt{d}\right).$$

*Proof.* Recall that every iteration defines $\alpha_i$ as $|G_i| - \nu \leq \alpha_i \leq |G_i|$. For every $i$, let $\hat{G}_i \subseteq G_i$ denote an arbitrary subset of size $\alpha_i$. Let $\hat{S}_e = \cup \hat{G}_i$.

We use the weak triangle inequality [14] stating that for every $x, y \in \mathbb{R}^d$ and a closed set $C \subseteq \mathbb{R}^d$ we have

$$|\text{dist}^2(x, C) - \text{dist}^2(y, C)| \leq \frac{12\|x - y\|_2^2}{\lambda} + \frac{\lambda}{2} \text{dist}^2(x, C).$$

In our case, by letting $x'$ denote the associated $c_i$ to $x \in \hat{S}_e$ when $x$ is deleted, and using $\lambda = 1/2$, this implies

$$
\begin{aligned}
|\text{cost}(\hat{S}_e, C) - \text{cost}(B, C)| &= \left| \sum_{x \in \hat{S}_e} \text{dist}^2(x, C) - \sum_{x \in \hat{S}_e} \text{dist}^2(x', C) \right| \\
&\leq \sum_{x \in \hat{S}_e} \left( \frac{12\|x - x'\|_2^2}{\lambda} + \frac{\lambda}{2} \text{dist}^2(x, C) \right) \\
&= 24 \sum_{i=1}^{|B|} \text{cost}(\hat{G}_i, c_i) + \frac{1}{4} \text{cost}(\hat{S}_e, C).
\end{aligned}
\tag{12}
$$

Hence,

$$\frac{3 \text{cost}(\hat{S}_e, C)}{4} \leq \text{cost}(B, C) + 24 \sum_{i=1}^{|B|} \text{cost}(\hat{G}_i, c_i).$$

By the fact that $C$ is an $O(1)$-approximation to the $k$-means of $B$, we get

$$\text{cost}(B, C) \leq O(1) \cdot \text{cost}(B, C^*).$$

Similarly to (12),

$$|\text{cost}(\hat{S}_e, C^*) - \text{cost}(B, C^*)| \leq 24 \sum_{i=1}^{|B|} \text{cost}(\hat{G}_i, c_i) + \frac{1}{4} \text{cost}(\hat{S}_e, C^*).$$

Hence,

$$\text{cost}(B, C^*) \leq \frac{5}{4} \text{cost}(\hat{S}_e, C^*) + 24 \sum_{i=1}^{|B|} \text{cost}(\hat{G}_i, c_i).$$

So,

$$
\begin{aligned}
\text{cost}(\hat{S}_e, C) &= O\left( \text{cost}(B, C) + \sum_{i=1}^{|B|} \text{cost}(\hat{G}_i, c_i) \right) \\
&= O\left( \text{cost}(B, C^*) + \sum_{i=1}^{|B|} \text{cost}(\hat{G}_i, c_i) \right) \\
&= O\left( \text{cost}(B, C^*) + \sum_{i=1}^{|B|} \text{cost}(G_i, c_i) \right) \\
&= O\left( \text{cost}(\hat{S}_e, C^*) + w^2 k \cdot \text{cost}(S, C^*) \right) \\
&= O(w^2 k) \cdot \text{cost}(S, C^*).
\end{aligned}
$$

Notice that $\text{cost}(S, C) = \text{cost}(\hat{S}_e, C) + \text{cost}(S \setminus \hat{S}_e, C)$, where $\left| S \setminus \hat{S}_e \right| \leq \nu \cdot |B| + t_{min}$. As there are at most $O(k \cdot \log(n))$ iterations (i.e., $|B| = O(k \cdot \log(n))$), we have that

$$\begin{aligned}
\text{cost}(S, C) &= \text{cost}(\hat{S}_e, C) + \text{cost}(S \setminus \hat{S}_e, C) \\
&\leq O(w^2 k) \cdot \text{cost}(S, C^*) + O\left(\nu k \log(n) + t_{min}\right) \cdot \text{diam}(X^d).
\end{aligned}$$

$\square$